\documentclass[fleqn,usenatbib]{mnras}

\usepackage{newtxtext,newtxmath}

\usepackage[T1]{fontenc}

\DeclareRobustCommand{\VAN}[3]{#2}
\let\VANthebibliography\thebibliography
\def\thebibliography{\DeclareRobustCommand{\VAN}[3]{##3}\VANthebibliography}

\usepackage{graphicx}	
\usepackage{amsmath}	
\usepackage{longtable}
\usepackage{threeparttable}
\usepackage[dvipsnames]{xcolor}
\usepackage{caption}
\usepackage{lineno}
\usepackage{comment}

\newcommand{\ha}{\hbox{H$\alpha$}}
\newcommand{\hb}{\hbox{H$\beta$}}

\newcommand{\oiii}{\hbox{O\,{\sc iii}}}

\newcommand{\hii}{\hbox{H\,{\sc ii}}}

\newcommand{\beagle}{\textsc{Beagle}}

\newcommand{\jwst}{\textit{JWST}}
\newcommand{\hst}{\textit{HST}}

\newcommand{\gsim}{\gtrsim}
\newcommand{\lsim}{\lesssim}

\newcommand{\muv}{$M_{{\rm UV}}$}
\newcommand{\ruv}{$r_{e, {\rm UV}}$}

\title[Star Forming Complexes in $z\simeq 6-8$  Galaxies]{JWST/NIRCam Observations of Stars and HII Regions  in $z\simeq 6-8$  Galaxies: Properties of Star Forming Complexes on 150 pc Scales}

\author[Z. Chen et al.]{
Zuyi Chen,$^{1}$\thanks{E-mail: zychen@arizona.edu}
Daniel P. Stark,$^{1}$
Ryan Endsley,$^{1}$
Michael Topping,$^{1}$ 
Lily Whitler,$^{1}$
\newauthor{
St\'ephane Charlot$^{2}$ 
}
\\
\vspace{0in}\\
$^{1}$Steward Observatory, University of Arizona, 933 N Cherry Ave, Tucson, AZ 85721 USA\\
$^{2}$Sorbonne Universit\'e, UPMC-CNRS, UMR7095, Institut d'Astrophysique de Paris, F-75014, Paris, France\\
}

\date{Accepted XXX. Received YYY; in original form ZZZ}

\pubyear{--}

\begin{document}
\label{firstpage}
\pagerange{\pageref{firstpage}--\pageref{lastpage}}
\maketitle

\begin{abstract}
The onset of the {\it JWST}-era provides a much-improved opportunity to characterize the resolved structure of early star forming systems. 
Previous {\it Spitzer} observations of $z\gsim 6$ galaxies revealed the presence of old stars and luminous HII regions (via [OIII]+H$\beta$ emission), but the poor resolution stunted our ability to map their locations with respect to the star forming regions identified in the rest-UV. 
In this paper, we investigate the internal structure of 12 of the most luminous $z\simeq 6-8$ galaxies in the EGS field observed with recent {\it JWST}/NIRCam imaging. 
The systems appear clumpy in the rest-UV, with more than half of the light coming from $\simeq 10^7$ to 10$^{9}$ M$_\odot$ star forming complexes that are $\simeq 150$ - 480 pc in size.
The clumps tend to be dominated by young stars (median = 36 Myr), but we also find large variations in clump ages within individual galaxies. 
The [OIII]+H$\beta$ EW varies significantly across individual galaxies (reflecting differences in stellar and gas properties), but the HII regions largely track the UV-bright complexes.
Perhaps surprisingly, the rest-optical continuum is just as clumpy as the UV, and we do not find older (and redder) nuclear stellar components that were previously undetected or faint in the UV. 
The majority of the stellar mass in bright $6<z<8$ galaxies  appears to be contained in the $\gsim 150$ pc-scale clumpy star forming complexes, reflecting the very active phase of assembly that is common in reionization-era galaxies.

\end{abstract}




\section{Introduction}

Over the last two decades, the {\it Hubble Space Telescope (HST)} has conducted a series of deep near-infrared imaging campaigns, enabling the identification of large samples of star forming galaxies at $z\simeq 6-10$ (\citealt{Bouwens2015,Finkelstein2015,McLeod2016,Bouwens2022}; see \citealt{Robertson2022} for a review). 
At these redshifts, HST images are limited to the rest-frame ultraviolet, revealing the unobscured star forming component of these early star forming systems.
Observations at 3-5 $\mu$m with the {\it Spitzer Space Telescope} have extended our view  into the rest-frame optical, providing a more robust census of the total stellar content of early galaxies. 
By combining  {\it HST} and {\it Spitzer} photometry for large samples at $z\gsim 6$, the bulk integrated properties of the population have been established (see \citealt{Stark2016} for a review),  revealing relatively low stellar masses ($\lsim 10^{10}$ M$_\odot$), large specific star formation rates ($\gsim 5-10$ Gyr$^{-1}$) and blue UV continuum slopes ($\mathbf{\beta \approx -2}$) indicative of minimal dust reddening.  
Collectively these observations are consistent with reionization-era galaxies 
undergoing rising star formation histories, leading to rapid 
mass growth and typical light-weighted stellar ages of 50-200 Myr \citep[e.g.,][]{Laporte2021,Whitler2022,Stefanon2022}.

Meaningfully building on this picture of early galaxy assembly 
will require a resolved view of the stars, gas, and dust within 
$z\gsim 6$ galaxies. The internal structure of early star 
forming systems has long been mostly out of reach observationally, 
with a picture mostly  limited to the rest-frame UV. 
Moderate resolution  {\it HST}/WFC3 images (FWHM = 0.2 arcsec) often reveal multiple-component kpc-scale clumpy structures in massive and UV-luminous reionization era galaxies \citep[e.g.,][]{Bowler2017,Marrone2018,Matthee2017,Matthee2019}
. Separations between the individual clumps are often upwards of 5 kpc in individual UV-luminous systems at $z\simeq 7$ \citep[e.g.,][]{Bowler2017}. 
Gravitational lensing has extended these studies to intrinsically fainter galaxies in deep cluster fields imaged by {\it HST}. After correcting for magnification, these dwarf galaxies are commonly revealed to be star forming components of size $\simeq 100$ pc, with the most highly-magnified sources probing 10 pc scales similar to super star clusters \citep[e.g.,][]{Kawamata2015,Vanzella2019,Vanzella2020,Bouwens2022b}. These  galaxies appear smaller than expected from extrapolation of size-luminosity scaling relations \citep[e.g.,][]{Shibuya2015}, leading some to suggest that the lowest luminosity galaxies probed 
by {\it HST} may only host a single star forming complex 
\citep[e.g.,][]{Bouwens2022b}. If true, the rest-UV properties (i.e., UV slope, M$_{\rm{UV}}$) would potentially only be applicable to a small sub-region within a larger galaxy \citep[e.g.,][]{Zick2018}.

Observations of gas and dust in luminous $z\simeq 7-8$ galaxies with the Atacama Large Millimeter Array (ALMA) have begun to complement the rest-UV picture from {\it HST} \citep{Matthee2017,Smit2018,Hashimoto2018,Bowler2022,Bouwens2022c}. 
In some galaxies, the dust continuum  is offset from the UV emission, suggesting that there are heavily-obscured star forming components that are faint or undetected in existing {\it HST} imaging \citep[e.g.,][]{Bowler2022}. 
In other systems with ALMA data, the dust emission is co-spatial but more extended than the UV \citep{{Bowler2022}}, again revealing star forming regions not well-recovered in the {\it HST} imaging.
Meanwhile, ALMA detections of far-IR fine structure emission lines (i.e,. [OIII]$\lambda$88$\mu$m and [CII]$\lambda$158$\mu$m) are changing our view of galaxies thought to have low stellar masses ($\simeq 10^9$ M$_\odot$) based on UV to optical SEDs dominated by very young ($\lsim10$ Myr) stellar populations.
Recent works have revealed significant spatial variations of [OIII] emission line properties (tracing ionized gas) in $z\simeq 7$ galaxies, suggesting these galaxies could be comprised of several stellar populations at different ages \citep{Wong2022,Witstok2022,Akins2022}.
It has also been shown that these
systems can have orders of magnitude larger dynamical masses (i.e., 10$^{11}$ M$_\odot$), potentially accommodating an older ($\gsim 100$ Myr) stellar population which (together with a large gas fraction) dominates the mass but is outshined by a recent burst \citep{Topping2022}.

A  complete picture of the internal structure of $z\gsim 6$ galaxies will 
ultimately require higher resolution imaging at longer wavelengths where the spatial distribution of old stars can clearly be established. 
At lower redshifts ($z\simeq 1.5-2.5)$, the clumpy star forming structures tend to be located at the outskirts 
of an older and dustier stellar component; the derived stellar mass distribution in these systems is much less clumpy than the rest-frame UV \citep[e.g.,][]{Wuyts2012,Lee2013}. The poor resolution of {\it Spitzer} (FWHM $\approx$ 2 arcsec) has long stood in the way of such an investigation at $z\gsim 6$, and as such it is not clear whether the UV-emitting regions identified by {\it HST} are co-spatial with the older stellar populations, or whether they are embedded (or at the outskirts) of a larger structure of older stars and gas. 

The spatial resolution provided by {\it JWST} at 2-5 $\mu$m enables the 
first resolved view of the rest-optical continuum emission in reionization-era galaxies, allowing the distribution of young and old stellar populations to be mapped within early galaxies. In this paper, we use  NIRCam imaging of the EGS field to study the internal structure of 12 reionization-era galaxies. The data were taken as part of the CEERS Early Release Science observations (ERS:1345, PI: Finkelstein). The systems chosen for this study are bright $z\simeq 6-8$ galaxies (F200W= 24.9 to 26.6)  identified photometrically in \cite{Bouwens2015} via  {\it HST} imaging of the EGS field. We describe the sample selection, NIRCam imaging reduction and analysis in \S2. In \S3, 
we use the NIRCam imaging to create maps of UV and optical continuum emission across each galaxy, investigating whether 
the star forming complexes are offset from the older stellar population which may dominate the light at longer wavelengths. We also probe the star 
forming regions via maps of strong rest-optical emission lines ([OIII]+H$\beta$) created from NIRCam long wavelength filters which are dominated by line emission. For each galaxy, we characterize the size and physical properties of the star forming complexes (i.e., stellar mass, age, star formation rate surface density). Finally we 
discuss implications for the build-up of galaxies in the reionization era, comparing these structures to systems at lower redshifts.

Throughout this paper, we adopt a flat $\Lambda$CDM cosmology with $H_0$ = 70 km s$^{-1}$ Mpc$^{-1}$, $\Omega_m = 0.3$, and $\Omega_\Lambda = 0.7$.
All magnitudes are measured in the AB system \citep{Oke1983} unless otherwise stated.
The equivalent widths are calculated in the rest-frame with positive values for emission.

\section{Data \& Sample Analysis}

\begin{table*}
    \centering
    \caption{UV-bright galaxies at $z\simeq 6-8$ in the CEERS NIRCam imaging footprint of the EGS field. 
    Our final sample consists of 12 galaxies that were selected from {\it HST} imaging of the field by \protect\cite{Bouwens2015}. 
    The names of objects are taken from \protect\cite{Bouwens2015}. Photometry and photometric redshifts are from updated NIRCam imaging. }
    \begin{threeparttable}[t]
    \begin{tabular}{lcccccccc}
    \hline
     Name & RA & Dec. & $z_{\rm phot}$ & F150W & F200W & F356W & F410M \\
     & & & & (mag) & (mag) & (mag) & (mag)\\
    \hline
     EGSI-9136950177 & 214.806959 & 52.838203 & $5.26_{-0.01}^{+0.01}$  & $24.94_{-0.01}^{+0.01}$  & $24.88_{-0.01}^{+0.01}$  & $24.68_{-0.01}^{+0.01}$  & $24.36_{-0.01}^{+0.01}$  \\ [2pt]
     EGSZ-9338153359 & 214.890898 & 52.893217 & $6.12_{-0.20}^{+0.23}$  & $26.63_{-0.04}^{+0.04}$  & $26.37_{-0.02}^{+0.02}$  & $25.64_{-0.01}^{+0.01}$  & $26.89_{-0.08}^{+0.09}$  \\ [2pt]
     EGSZ-9314453285 & 214.881001 & 52.891207 & $6.54_{-0.09}^{+0.04}$  & $26.38_{-0.04}^{+0.04}$  & $26.09_{-0.03}^{+0.03}$  & $25.57_{-0.01}^{+0.01}$  & $26.35_{-0.05}^{+0.06}$  \\ [2pt]
     EGSI-9179549499 & 214.824915 & 52.830484 & $6.56_{-0.10}^{+0.04}$  & $25.95_{-0.03}^{+0.03}$  & $25.76_{-0.03}^{+0.03}$  & $25.53_{-0.01}^{+0.01}$  & $26.20_{-0.04}^{+0.04}$  \\ [2pt]
     EGSZ-9271353221 & 214.863024 & 52.889431 & $6.59_{-0.04}^{+0.03}$  & $26.10_{-0.03}^{+0.03}$  & $25.93_{-0.02}^{+0.02}$  & $25.44_{-0.02}^{+0.02}$  & $26.39_{-0.08}^{+0.08}$  \\ [2pt]
     EGSZ-9419055074 & 214.924558 & 52.918690 & $6.66_{-0.01}^{+0.01}$  & $26.04_{-0.05}^{+0.05}$  & $25.72_{-0.02}^{+0.02}$  & $24.51_{-0.01}^{+0.01}$  & $25.42_{-0.02}^{+0.03}$  \\ [2pt]
     EGSZ-9350655307 & 214.896097 & 52.925157 & $6.69_{-0.02}^{+0.02}$  & $25.96_{-0.03}^{+0.03}$  & $25.97_{-0.02}^{+0.02}$  & $25.26_{-0.01}^{+0.01}$  & $25.57_{-0.03}^{+0.03}$  \\ [2pt]
     EGSI-0020500266 & 215.008510 & 53.007336 & $6.74_{-0.03}^{+0.04}$  & $26.48_{-0.03}^{+0.03}$  & $26.44_{-0.03}^{+0.03}$  & $25.65_{-0.01}^{+0.01}$  & $25.68_{-0.02}^{+0.02}$  \\ [2pt]
     EGSZ-9135048459 & 214.806240 & 52.812713 & $6.74_{-0.01}^{+0.01}$  & $26.88_{-0.05}^{+0.05}$  & $27.25_{-0.05}^{+0.05}$  & $27.27_{-0.05}^{+0.05}$  & $27.40_{-0.09}^{+0.10}$  \\ [2pt]
     EGSZ-9262051131 & 214.859155 & 52.853586 & $7.25_{-0.11}^{+0.11}$  & $26.24_{-0.04}^{+0.04}$  & $26.23_{-0.05}^{+0.05}$  & $26.23_{-0.03}^{+0.03}$  & $25.22_{-0.02}^{+0.02}$  \\ [2pt]
     EGSY-9105550297 & 214.793938 & 52.841534 & $7.32_{-0.09}^{+0.23}$  & $25.60_{-0.02}^{+0.02}$  & $25.77_{-0.02}^{+0.02}$  & $25.58_{-0.01}^{+0.01}$  & $25.02_{-0.02}^{+0.02}$  \\ [2pt]
     EGSY-9587400281 & 214.994754 & 53.007741 & $7.54_{-0.04}^{+0.03}$  & $26.24_{-0.03}^{+0.03}$  & $26.08_{-0.02}^{+0.02}$  & $26.01_{-0.02}^{+0.02}$  & $25.23_{-0.02}^{+0.02}$  \\ [2pt]
     \hline
    \end{tabular}
    \end{threeparttable}
    \label{tab:sample}
\end{table*}

In this section, we first describe the NIRCam imaging used for 
our analysis (\S2.1), before discussing the sample selection and 
integrated galaxy properties (\S2.2) and our spatially-resolved 
analysis (\S2.3). 

\subsection{NIRCam imaging}

The \jwst{}/NIRCam imaging utilized in this work were taken as part of the Cosmic Evolution Early Release Science \citep[CEERS\footnote{\url{https://ceers.github.io/}}, ERS:1345;][]{Finkelstein2017} survey during June 2022 (see \citealt{Bagley2022} for a more detailed description of the observations). 
These NIRCam data include imaging in six broad-band filters (F115W, F150W, F200W, F277W, F356W, and F444W) as well as one medium-band filter (F410M) across four independent pointings for a total on-sky coverage of $\approx$40 arcmin$^2$.
Imaging in the short wavelength (SW) filters (F115W, F150W, and F200W) and long wavelength (LW) filters (F277W, F356W, F444W, and F410M) have a resolution of 0.062\arcsec{} and 0.125\arcsec{}, respectively.
At $z\approx6 - 8$, these filters simultaneously probe the rest-frame UV light dominated by young massive stars and the optical emission from the nebular and stellar continuum as well as strong emission lines (e.g., \hb{}, [\oiii{}], and \ha{}).
This enables us to resolve reionization-era galaxies in the rest UV and optical at comparably high resolution.
While a detailed description of the NIRCam imaging data reduction can be found in \cite{Endsley2022}, we briefly summarize the major steps below.

To produce co-added mosaics for each NIRCam filter, we first retrieve the calibrated flat-fielded individual exposures ({\tt *\_cal.fits}) from the MAST Portal\footnote{\url{https://mast.stsci.edu/portal/Mashup/Clients/Mast/Portal.html}} and subsequently perform a global background subtraction on each science exposure using the \textsc{sep} package\footnote{\url{https://sep.readthedocs.io/en/v1.0.x/index.html}}.
To provide the best possible astrometric alignment, we next process images through {\tt calwebb\_image3}\footnote{\url{https://jwst-pipeline.readthedocs.io/en/latest/index.html}} separately for each NIRCam filter, pointing, and module combination given a slight ($\sim$0.2 arcsec) offset in the world coordinate system of the NIRCam module A vs. B fits file headers (private communication with NIRCam instrument team).
In this process, we have corrected the $1/f$ noise, obtained robust background estimates, and implemented updated photometric calibrations specific to each NIRCam detector derived by Gabe Brammer\footnote{\url{https://github.com/gbrammer/grizli/pull/107}} (see \citealt{Endsley2022}).
Each of the output co-adds ({\tt *\_i2d.fits}) are then separately aligned to the \textit{Gaia} astrometric frame using the \textsc{tweakreg} package\footnote{\url{https://drizzlepac.readthedocs.io/en/latest/tweakreg.html}}, adopting the Gaia-aligned \hst{}/WFC3 F160W mosaic from CHArGE (see below) as a reference image.
It was not possible to perform direct astrometric alignment to the \textit{Gaia} frame for each co-added NIRCam image due to the low density of unsaturated \textit{Gaia} stars with proper motion measurements in the EGS field ($\lesssim$3 per pointing and module combination).
Nonetheless, with \textsc{tweakreg} we are able to achieve sub-pixel accuracy of the resulting astrometry, with an RMS offset of $\approx$6--15 mas relative to the \hst{} CHArGE mosaics.
We then combine all co-added images for each filter into a single mosaic by resampling them onto a grid with a pixel scale of 30 mas pixel$^{-1}$, where the world coordinate system of this grid is the same for all filters.

We also use \hst{}/ACS imaging in the EGS field for constraining the physical properties of the galaxies analyzed in this work. 
The ACS mosaics utilized here were produced as part of the Complete Hubble Archive for Galaxy Evolution (CHArGE) project (Kokorev in prep.) and are  matched to the \textit{Gaia} astrometric frame with a pixel scale of 40 mas pixel$^{-1}$.
These CHArGE mosaics include data in the F435W, F606W, and F814W bands as part of the following surveys: the All-Wavelength Extended Groth Strip International Survey (AEGIS; \citealt{Davis2007}), the Cosmic Assembly Near-infrared Deep Extragalactic Legacy Survey (CANDELS; \citealt{Grogin2011,Koekemoer2011}), and the Ultraviolet Imaging of the Cosmic Assembly Near-infrared Deep Extragalactic Legacy Survey Fields (UVCANDELS\footnote{\url{https://archive.stsci.edu/hlsp/uvcandels}}; PI: Teplitz).

The \hst{} and \jwst{} imaging introduced above span nearly a factor of four in angular resolution from a PSF full width at half maximum (FWHM) of $\approx$0.045 arcsec in ACS/F435W to $\approx$0.18 arcsec in WFC3/F160W.
To ensure robust color calculations, we must therefore account for variations in the PSF across different images.
At the same time, we aim to preserve as much signal-to-noise as possible, particularly in the shorter wavelength filters which probe the Lyman-alpha break.
We have thus opted to homogenize the PSFs of all the NIRCam LW mosaics (as well as the WFC3 mosaics used by our companion papers) to that of WFC3/F160W, while the ACS and NIRCam SW mosaics are homogenized to the PSF of ACS/F814W.\footnote{The ACS/F814W and WFC3/F160W mosaics have slightly larger FWHM relative to that of NIRCam/F200W and NIRCam/F444W, respectively, given the smaller mirror size of \hst{}.}
Our algorithm for performing the PSF homogenization largely follows that of \citet{Endsley2021} where we construct the PSF of each mosaic by stacking cutouts of unsaturated isolated stars (oversampling to a resolution of 3 mas pixel$^{-1}$ for this step) and then derive PSF convolution kernels using the Python \textsc{photutils} package \citep{Bradley2021}.
The PSF encircled energy distributions after convolution are found to agree with the targeted PSF (i.e. ACS/F814W or WFC3/F160W) to within 1--2 per cent at radii $>$0.1\arcsec{} in all bands.

\subsection{Sample Selection and Galaxy  Properties}\label{sec:sample}

\begin{figure*}
    \centering
    \includegraphics[width=1.0\textwidth]{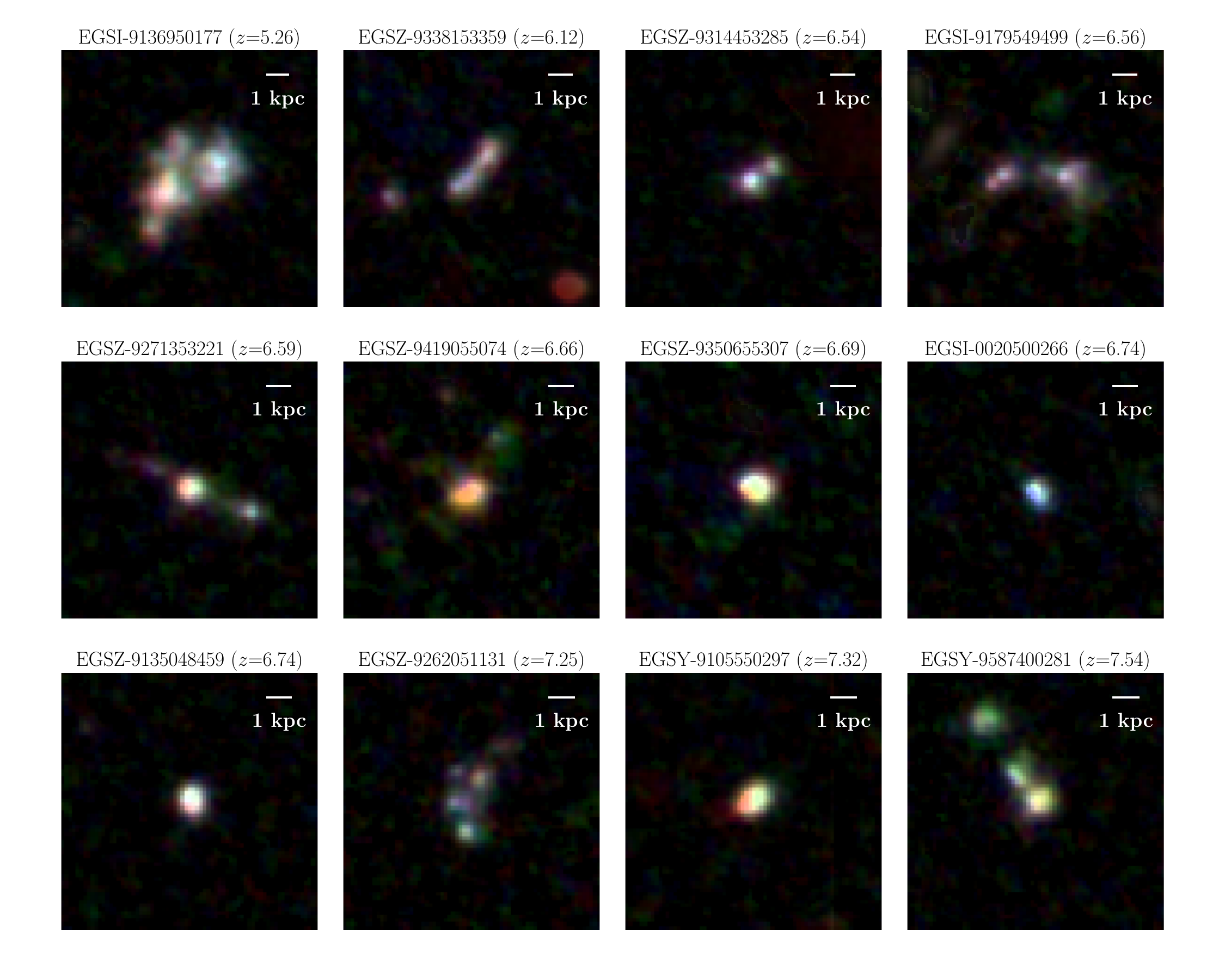}
    \caption{\jwst{}/NIRCam color images of the 12 $z\simeq 6-8$ galaxies analyzed in this paper. {\em Blue}: NIRCam/F115W, {\em green}: NIRCam/F150W, {\em red}: NIRCam/F200W.
    The images span a scale of 2\arcsec{}x2\arcsec{}.
    The nearby foreground objects in each image (identified with clear detections in the \hst{} optical filters) have been masked. These galaxies were selected in the {\it HST} imaging of the EGS and are among the brightest star forming galaxies (F200W=24.8-26.6) in the CEERS footprint at $z\gsim 6$. The filters correspond to the rest-frame UV. The  light is dominated by a number of clumpy star forming complexes, with sizes ranging from unresolved ($<$150 pc) to 480 pc.   }
    \label{fig:rgb}
\end{figure*}

\begin{figure*}
    \centering
    \includegraphics[width=1\textwidth]{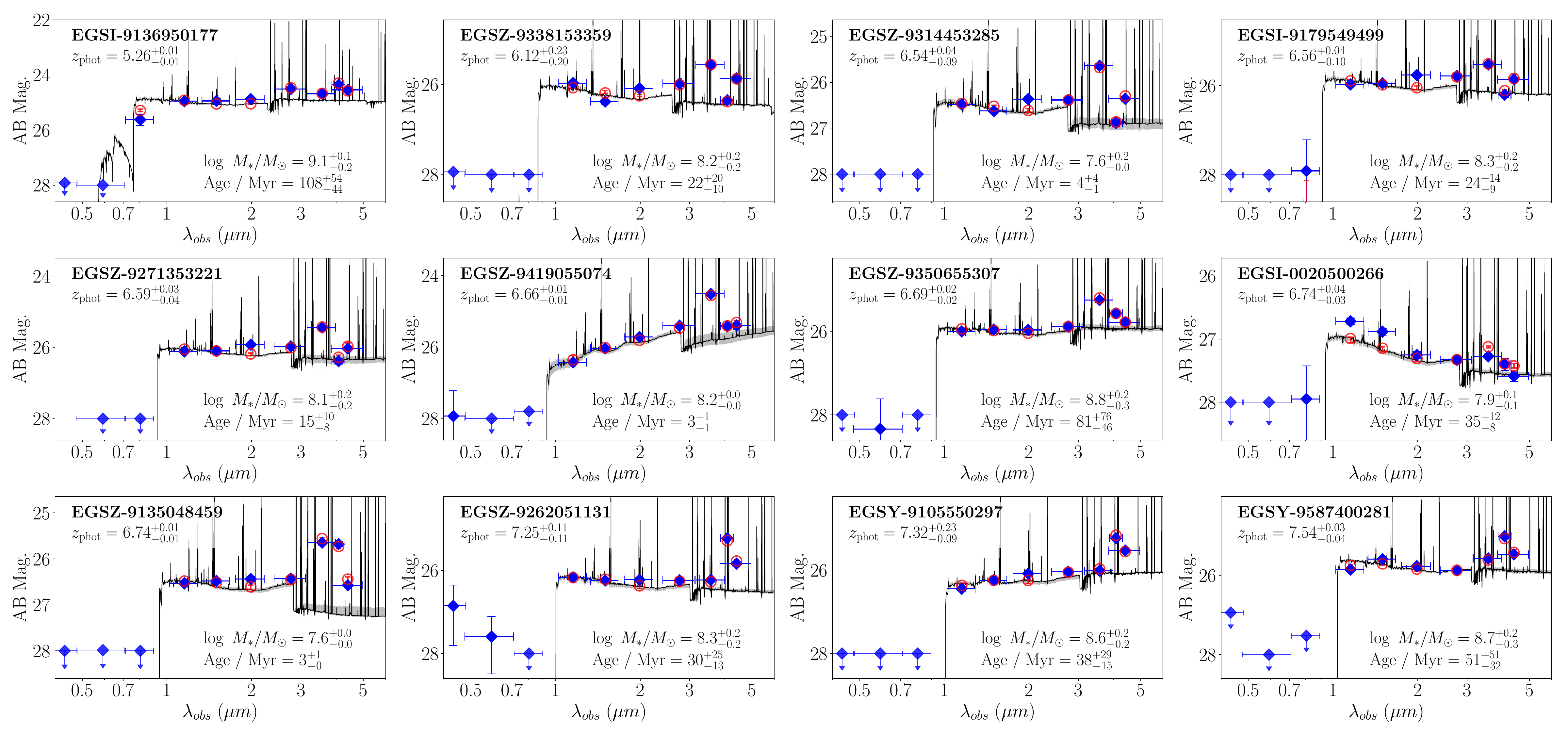}
    \caption{The \beagle{} SED fits to the 12 $z\simeq 6-8$ UV-bright galaxies analyzed in this paper. 
    The optical data from {\it HST}/ACS and near-infrared measurements from {\it JWST}/NIRCam are marked with blue diamonds, and we show  $2\sigma$ upper limits shown in cases of non-detections.
    Overlaid on the data are the fitted median (black line) and 68 per cent credible interval SEDs (grey shaded regions) from \beagle{}.
    The corresponding model photometry is also shown in red circles.
    The stellar mass and constant star formation history age are shown for each object. The objects have light-weighted ages that range between 3 Myr and 183 Myr, with stellar masses between 4$\times$10$^7$ M$_\odot$ and 2$\times$10$^9$ M$_\odot$.}
    \label{fig:beagle}
\end{figure*}

\begin{table*}
    \centering
    \caption{Physical properties inferred from {\sc BEAGLE} by SED-fitting of \hst{}/ACS+\jwst{}/NIRCam photometry.
    We have adopted a constant star formation history.
    The fitted photometric redshifts have been presented in Table \ref{tab:sample}.
    Here, we report the absolute rest-UV magnitude, the stellar mass, the specific star formation rates and the stellar population ages (defined as the time since star formation starts).
    We also list their $V$-band optical depth and the [\oiii{}]+\hb{} emission line equivalent widths.
    The reported values are the median of the posterior with the 16th and 84th percentiles as uncertainties.}
    \begin{tabular}{ccccccccc}
    \hline
         Name & $M_{\rm UV}$ & log ($M_*/M_\odot$) & sSFR & Age & $\tau_{\scriptscriptstyle{V}}$ & EW [\oiii{}]+\hb{}\\
         & (mag) & & (Gyr$^{-1}$) & (Myr) & & (\AA{})\\
    \hline
EGSI-9136950177       & $-21.60_{-0.03}^{+0.06}$  & $9.13_{-0.17}^{+0.12}$  & $9_{-2}^{+6}$  & $108_{-44}^{+54}$  & $0.01_{-0.01}^{+0.02}$  & $773_{-31}^{+223}$  \\ [2pt]
EGSZ-9338153359       & $-20.31_{-0.03}^{+0.07}$  & $7.60_{-0.04}^{+0.18}$  & $44_{-21}^{+35}$  & $4_{-1}^{+4}$  & $0.01_{-0.01}^{+0.02}$  & $2554_{-193}^{+195}$  \\ [2pt]
EGSZ-9314453285       & $-20.67_{-0.02}^{+0.08}$  & $8.18_{-0.21}^{+0.23}$  & $236_{-111}^{+65}$  & $22_{-10}^{+20}$  & $0.01_{-0.00}^{+0.01}$  & $1387_{-116}^{+199}$  \\ [2pt]
EGSI-9179549499       & $-20.95_{-0.04}^{+0.06}$  & $8.30_{-0.16}^{+0.16}$  & $41_{-15}^{+24}$  & $24_{-9}^{+14}$  & $0.01_{-0.01}^{+0.02}$  & $990_{-95}^{+90}$  \\ [2pt]
EGSZ-9271353221       & $-20.78_{-0.02}^{+0.04}$  & $8.11_{-0.25}^{+0.18}$  & $66_{-26}^{+76}$  & $15_{-8}^{+10}$  & $0.01_{-0.01}^{+0.02}$  & $1613_{-318}^{+209}$  \\ [2pt]
EGSZ-9419055074       & $-20.40_{-0.03}^{+0.02}$  & $8.20_{-0.03}^{+0.05}$  & $337_{-88}^{+79}$  & $3_{-1}^{+1}$  & $0.31_{-0.03}^{+0.03}$  & $2130_{-226}^{+386}$  \\ [2pt]
EGSZ-9350655307       & $-20.91_{-0.04}^{+0.07}$  & $8.81_{-0.26}^{+0.19}$  & $12_{-6}^{+14}$  & $81_{-46}^{+76}$  & $0.05_{-0.04}^{+0.03}$  & $772_{-158}^{+261}$  \\ [2pt]
EGSZ-9135048459       & $-19.86_{-0.07}^{+0.06}$  & $7.93_{-0.09}^{+0.09}$  & $361_{-58}^{+55}$  & $35_{-8}^{+12}$  & $0.00_{-0.00}^{+0.00}$  & $456_{-52}^{+121}$  \\ [2pt]
EGSI-0020500266       & $-20.40_{-0.03}^{+0.03}$  & $7.57_{-0.02}^{+0.02}$  & $28_{-7}^{+8}$  & $3_{-0}^{+1}$  & $0.01_{-0.01}^{+0.01}$  & $3436_{-645}^{+356}$  \\ [2pt]
EGSZ-9262051131       & $-20.84_{-0.05}^{+0.09}$  & $8.35_{-0.19}^{+0.20}$  & $31_{-13}^{+23}$  & $30_{-13}^{+25}$  & $0.01_{-0.00}^{+0.01}$  & $1037_{-200}^{+21}$  \\ [2pt]
EGSY-9105550297       & $-21.36_{-0.02}^{+0.05}$  & $8.72_{-0.30}^{+0.23}$  & $26_{-11}^{+18}$  & $51_{-32}^{+51}$  & $0.01_{-0.01}^{+0.02}$  & $841_{-114}^{+73}$  \\ [2pt]
EGSY-9587400281       & $-20.72_{-0.04}^{+0.05}$  & $8.59_{-0.17}^{+0.18}$  & $19_{-9}^{+31}$  & $38_{-15}^{+29}$  & $0.14_{-0.03}^{+0.02}$  & $590_{-52}^{+64}$  \\ [2pt]
    \hline
    \end{tabular}
    \label{tab:beagle}
\end{table*}

The goal for this study is to provide a resolved view of the brightest galaxies in the reionization era.
Our sample is based on the galaxies previously identified in existing \hst{} imaging of the EGS field \citep{Bouwens2015,Roberts-Borsani2016,Bouwens2016,Bouwens2019,Leonova2022}.
We limit our selection to those systems that fall into the CEERS/NIRCam footprint observed on June 29, 2022.
Our current study is focused on 12 galaxies with photometric redshifts between $6<z<8$ with H-band magnitudes brighter than {\it H} = 26.6 mag to ensure that we are able to resolve enough pixels at the required signal-to-noise ratio with the NIRCam data.
However, we note that in a future paper, we will investigate the resolved properties of the more general population of high-z galaxies with a well-selected sample directly from the NIRCam imaging.
We list the 12 \hst{}-selected bright galaxies in Table \ref{tab:sample}, and adopted the source names from the original literature sources.
Color postage stamp images based on the  NIRCam SW imaging  are also presented in Figure \ref{fig:rgb}. 

The NIRCam photometry in the seven filters for our sample are extracted with custom elliptical apertures.
In this process, we ensure the aperture includes the total galaxy flux as indicated by {\sc Photutils} \citep{Bradley2021} segmentation maps in all filters, while also avoiding light from neighbouring contaminants.
These contaminants are identified by their detections in the \hst{}/ACS optical filters, suggesting they lie at lower redshifts than our galaxies.
Photometry uncertainties are estimated by placing the same aperture in the surrounding 10\arcsec{}x10\arcsec{} regions with all the sources being masked.
The photometry is then corrected for aperture loss, which is estimated by centering the aperture at the corresponding PSF.
We list the NIRCam photometry in four of the filters (F150W and F200W sampling the rest-UV,  and F356W and F444W sampling the rest-optical) in Table \ref{tab:sample}.
In the same manner, we also extract the \hst{}/ACS short-wavelength filters (F435W, F606W, and F814W) photometry, which will be used for constraining the photometric redshift when deriving galaxy properties.

Following the procedure described in \cite{Whitler2022}, we fit the \hst{}+NIRcam photometry with BayEsian
Analysis of GaLaxy sEds \citep[{\sc BEAGLE v0.20.4};][]{Chevallard2016}, providing constraints on the physical properties of our galaxies.
The \cite{Gutkin2016} models are adopted, which self-consistently combine the latest version of \cite{Bruzual2003} stellar population models underpinned by the {\sc parsec} isochrones \citep{Bressan2012,Chen2015} and the nebular emission computed by {\sc Cloudy} \citep{Ferland2013}. 
We place uniform prior on redshift from $5\leq z \leq 10$, which allows us to obtain updated constraints on the photometric redshifts with the inclusion of NIRCam photometry.
A constant star formation history (CSFH) is assumed for our fiducial models, with the total stellar mass allowed to vary from $5\leq {\rm log}(M_*/M_\odot) \leq 12$ and the star formation period from 1 Myr to the age of the Universe at the corresponding redshift. We note that CSFH models can underpredict the stellar mass and age of galaxies dominated by 
very young light-weighted ages \citep[e.g.,][]{Tang2022,Whitler2022,Tacchella2022}, and we thus also consider more flexible star formation histories in our analysis later in this paper. 
We adopt the \cite{Chabrier2003} initial mass function with the upper-mass cutoff of 300 $M_\odot$.
Narrow log-normal priors are used on stellar metallicity (center $\mu_{{\rm log} (Z/Z_\odot)} = -0.7$, standard deviation $\sigma_{{\rm log} (Z/Z_\odot)} = 0.15$) and ionization parameter ($\mu_{{\rm log} U} = -2.5$, $\sigma_{{\rm log} U} = 0.25$), as motivated by spectroscopic observations of the high-ionization emission lines at $z$ = 7 -- 9 \citep[e.g.,][]{Stark2017,Hutchison2019}.
The nebular metallicity (including dust and gas-phase) is kept the same as the stellar metallicity, with a fixed dust-to-metal mass ratio of $\xi_{\rm d} = 0.3$. Future investigation will investigate the impact of alpha enhancement, but we note that the effects of depletion do decouple the gas and stellar metallicity in our models. Finally, we assume the SMC dust attenuation curve \citep{Pei1992} with V-band optical depth varying from $-3.0\leq {\rm log}(\tau_{\rm \scriptscriptstyle{V}})\leq0.7$, and adopt the \cite{Inoue2014} model for the intergalactic medium (IGM) attenuation.

The spectral energy distribution (SED) fits are shown in Figure \ref{fig:beagle}.  The resulting constraints on the integrated properties, such as absolute UV magnitude (M$_{\scriptscriptstyle{UV}}$), stellar mass ($M_*$), and the specific star formation rates (sSFRs), are summarized in Table \ref{tab:beagle}.
The updated photometric redshifts derived from the new NIRCam photometry (together with {\it HST}/ACS constraints) place our galaxies at $5.26 \leq z \leq 7.54$.
The \muv{} varies from -19.86 to -21.60, with the median absolute magnitude (-20.8) near  M$_{\scriptscriptstyle{UV}}^*$, assuming the $z\sim 7$ luminosity function of \citealt{Bowler2017}.
The  galaxies in our sample have moderate stellar masses that range from $3.7\times10^7$ $M_\odot$ to $1.3\times10^9$ $M_\odot$.  Adoption of more flexible star formation histories can increases these masses by a factor of several for the systems with the youngest light-weighted ages (see \citealt{Whitler2022,Topping2022}).
The galaxies are found to span a range of sSFRs (9 -- 361 Gyr$^{-1}$) and light-weighted stellar population ages (3 to 107 Myr), both derived assuming CSFH.
The youngest among them are likely experiencing a recent upturn or burst of star formation, powering high [\oiii{}]+\hb{} equivalent widths (456 -- 3436 \AA{}).
These emission lines can heavily contaminate the NIRCam filters, allowing us to map the spatial distribution of these lines in the following sections.

\begin{table*}
    \centering
    \caption{The physical properties of the individual clumps in each galaxy.
    Clumps within the same galaxy are labeled and ordered by decreasing apparent F200W magnitudes. 
    The rest-optical sizes are left blank when the long-wavelength NIRCam resolution does not allow us to resolve two smaller structures identified in the UV.}
    \begin{threeparttable}[t]
    \begin{tabular}{cccccccc}
    \hline
         Name & \muv{} & $r_{e,{\rm UV}}$ & $r_{e,{\rm opt}}$ & log ($M_*/M_\odot$) & Age & $\tau_{\scriptscriptstyle{V}}$ & EW [\oiii{}]+\hb{} \\
         & (mag) & (pc) & (pc) & &(Myr) & & (\AA{})\\
    \hline
9136-C1     &  $-20.17_{-0.02}^{+0.02}$ &  $<       367$ &  $<       735$ &  $8.74_{-0.16}^{+0.12}$ &  $173_{- 71}^{+ 88}$ &  $0.02_{-0.01}^{+0.03}$ &  $ 817_{-  18}^{+ 182}$ \\[2pt]
9136-C2     &  $-20.05_{-0.04}^{+0.06}$ &  $<       367$ &  $<       735$ &  $8.17_{-0.10}^{+0.10}$ &  $ 48_{- 12}^{+ 17}$ &  $0.00_{-0.00}^{+0.00}$ &  $1003_{- 114}^{+ 162}$ \\[2pt]
9136-C3     &  $-19.42_{-0.02}^{+0.05}$ &  $<       367$ &   --           &  $8.23_{-0.20}^{+0.13}$ &  $101_{- 46}^{+ 53}$ &  $0.01_{-0.00}^{+0.02}$ &  $ 949_{-  87}^{+  88}$ \\[2pt]
9136-C4     &  $-18.89_{-0.09}^{+0.05}$ &  $<       367$ &  $<       367$ &  $8.15_{-0.28}^{+0.23}$ &  $ 89_{- 51}^{+110}$ &  $0.09_{-0.04}^{+0.03}$ &  $1063_{- 266}^{+ 101}$ \\[2pt]
\hline

9338-C1     &  $-19.65_{-0.04}^{+0.05}$ &  $ 387\pm  39$ &  $ 356\pm  13$ &  $8.66_{-0.11}^{+0.09}$ &  $260_{- 92}^{+ 85}$ &  $0.01_{-0.01}^{+0.04}$ &  $ 505_{-  21}^{+  16}$ \\[2pt]
9338-C2     &  $-19.45_{-0.06}^{+0.08}$ &  $<       169$ &  --            &  $7.78_{-0.14}^{+0.14}$ &  $ 32_{- 11}^{+ 18}$ &  $0.00_{-0.00}^{+0.01}$ &  $1000_{- 128}^{+ 159}$ \\[2pt]
9338-C3     &  $-19.27_{-0.02}^{+0.08}$ &  $<       169$ &  --            &  $7.61_{-0.10}^{+0.11}$ &  $ 25_{-  6}^{+ 10}$ &  $0.00_{-0.00}^{+0.00}$ &  $1129_{- 119}^{+  91}$ \\[2pt]
9338-C4     &  $-18.80_{-0.06}^{+0.11}$ &  $ 409\pm  45$ &  $<       339$ &  $6.94_{-0.02}^{+0.02}$ &  $  2_{-  0}^{+  0}$ &  $0.00_{-0.00}^{+0.01}$ &  $3841_{- 166}^{+ 422}$ \\[2pt]
\hline

9314-C1     &  $-19.96_{-0.03}^{+0.02}$ &  $ 285\pm  13$ &  $ 416\pm  22$ &  $7.38_{-0.02}^{+0.42}$ &  $  2_{-  0}^{+ 15}$ &  $0.00_{-0.00}^{+0.00}$ &  $1611_{- 173}^{+ 631}$ \\[2pt]
9314-C2     &  $-19.41_{-0.03}^{+0.04}$ &  $ 194\pm  14$ &  $<       326$ &  $7.29_{-0.03}^{+0.05}$ &  $  3_{-  1}^{+  1}$ &  $0.06_{-0.03}^{+0.03}$ &  $2971_{- 381}^{+ 242}$ \\[2pt]
\hline

9179-C1     &  $-19.86_{-0.06}^{+0.03}$ &  $ 391\pm  92$ &  $ 627\pm  84$ &  $8.12_{-0.12}^{+0.11}$ &  $ 54_{- 16}^{+ 22}$ &  $0.00_{-0.00}^{+0.01}$ &  $ 735_{- 143}^{+  66}$ \\[2pt]
9179-C2     &  $-19.63_{-0.01}^{+0.01}$ &  $ 330\pm  52$ &  $<       325$ &  $7.60_{-0.20}^{+0.20}$ &  $ 14_{-  6}^{+ 10}$ &  $0.01_{-0.00}^{+0.01}$ &  $1504_{- 111}^{+ 173}$ \\[2pt]
9179-C3     &  $-19.09_{-0.06}^{+0.08}$ &  $<       162$ &   --          &  $7.17_{-0.06}^{+0.26}$ &  $  5_{-  2}^{+  6}$ &  $0.04_{-0.03}^{+0.03}$ &  $2023_{-  34}^{+  44}$ \\[2pt]
\hline

9271-C1     &  $-20.15_{-0.02}^{+0.10}$ &  $ 256\pm  16$ &  $ 365\pm  25$ &  $7.60_{-0.03}^{+0.04}$ &  $  3_{-  0}^{+  1}$ &  $0.09_{-0.03}^{+0.02}$ &  $2635_{- 283}^{+ 343}$ \\[2pt]
9271-C2     &  $-19.11_{-0.03}^{+0.03}$ &  $ 350\pm  32$ &  $ 399\pm  29$ &  $7.69_{-0.10}^{+0.10}$ &  $ 39_{- 10}^{+ 14}$ &  $0.00_{-0.00}^{+0.00}$ &  $ 762_{- 194}^{+  94}$ \\[2pt]
9271-C3     &  $-18.66_{-0.02}^{+0.06}$ &  $ 271\pm  48$ &  $<       324$ &  $7.57_{-0.29}^{+0.32}$ &  $ 39_{- 23}^{+ 55}$ &  $0.01_{-0.01}^{+0.03}$ &  $1125_{- 304}^{+  98}$ \\[2pt]
\hline

9419-C1     &  $-20.05_{-0.02}^{+0.06}$ &  $ 479\pm  51$ &  $ 541\pm  67$ &  $8.05_{-0.02}^{+0.05}$ &  $  4_{-  1}^{+  2}$ &  $0.31_{-0.02}^{+0.02}$ &  $2272_{- 476}^{+  66}$ \\[2pt]
9419-C2     &  $-18.32_{-0.05}^{+0.02}$ &  $ 222\pm  35$ &  $<       323$ &  $7.82_{-0.23}^{+0.18}$ &  $115_{- 58}^{+ 86}$ &  $0.01_{-0.00}^{+0.02}$ &  $ 658_{- 219}^{+ 203}$ \\[2pt]
\hline

9350-C1     &  $-20.67_{-0.03}^{+0.04}$ &  $ 251\pm  25$ &  $<       322$ &  $8.74_{-0.18}^{+0.17}$ &  $ 73_{- 33}^{+ 57}$ &  $0.10_{-0.03}^{+0.03}$ &  $ 873_{- 133}^{+  90}$ \\[2pt]
\hline

0020-C1     &  $-19.84_{-0.01}^{+0.01}$ &  $<       160$ &  $<       320$ &  $7.87_{-0.08}^{+0.08}$ &  $ 32_{-  7}^{+  8}$ &  $0.00_{-0.00}^{+0.00}$ &  $ 319_{-  23}^{+  72}$ \\[2pt]
\hline

9135-C1     &  $-20.32_{-0.02}^{+0.01}$ &  $<       160$ &  $<       320$ &  $7.57_{-0.03}^{+0.04}$ &  $  2_{-  1}^{+  0}$ &  $0.00_{-0.00}^{+0.01}$ &  $3777_{- 307}^{+ 448}$ \\[2pt]
\hline

9262-C1     &  $-19.38_{-0.04}^{+0.03}$ &  $<       153$ &  $<       307$ &  $8.22_{-0.23}^{+0.18}$ &  $ 95_{- 49}^{+ 77}$ &  $0.05_{-0.04}^{+0.03}$ &  $ 680_{- 132}^{+  51}$ \\[2pt]
9262-C2     &  $-19.46_{-0.05}^{+0.07}$ &  $<       153$ &  $<       307$ &  $7.83_{-0.13}^{+0.14}$ &  $ 42_{- 13}^{+ 21}$ &  $0.00_{-0.00}^{+0.00}$ &  $ 539_{-  88}^{+ 227}$ \\[2pt]
9262-C3     &  $-19.42_{-0.04}^{+0.07}$ &  $<       153$ &  $<       307$ &  $7.69_{-0.11}^{+0.13}$ &  $ 24_{-  7}^{+ 11}$ &  $0.00_{-0.00}^{+0.00}$ &  $1277_{- 112}^{+  31}$ \\[2pt]
9262-C4     &  $-18.89_{-0.03}^{+0.01}$ &  $<       153$ &  --           &  $7.57_{-0.25}^{+0.25}$ &  $ 27_{- 14}^{+ 27}$ &  $0.01_{-0.01}^{+0.02}$ &  $1232_{- 364}^{+ 315}$ \\[2pt]
\hline

9105-C1     &  $-20.50_{-0.01}^{+0.01}$ &  $ 248\pm   7$ &  $ 364\pm  35$ &  $8.56_{-0.16}^{+0.15}$ &  $ 51_{- 20}^{+ 33}$ &  $0.12_{-0.03}^{+0.02}$ &  $ 552_{-  95}^{+ 134}$ \\[2pt]
\hline

9587-C1     &  $-20.59_{-0.03}^{+0.01}$ &  $ 377\pm  14$ &  $ 587\pm  23$ &  $8.36_{-0.23}^{+0.20}$ &  $ 42_{- 21}^{+ 37}$ &  $0.03_{-0.02}^{+0.02}$ &  $ 576_{- 144}^{+  64}$ \\[2pt]
9587-C2     &  $-19.93_{-0.00}^{+0.05}$ &  $<       149$ &  --           &  $7.91_{-0.12}^{+0.15}$ &  $ 33_{- 10}^{+ 16}$ &  $0.00_{-0.00}^{+0.00}$ &  $ 527_{-  13}^{+ 146}$ \\[2pt]
9587-C3     &  $-19.76_{-0.02}^{+0.00}$ &  $ 374\pm  22$ &  $ 506\pm  23$ &  $8.27_{-0.20}^{+0.18}$ &  $ 91_{- 43}^{+ 68}$ &  $0.01_{-0.00}^{+0.01}$ &  $ 564_{- 130}^{+ 133}$ \\[2pt]
9587-C4     &  $-19.66_{-0.04}^{+0.08}$ &  $<       149$ &  --           &  $7.81_{-0.11}^{+0.13}$ &  $ 31_{-  8}^{+ 15}$ &  $0.00_{-0.00}^{+0.00}$ &  $ 665_{-  18}^{+ 163}$ \\[2pt]
\hline
\hline
    \end{tabular}
    \end{threeparttable}
    \label{tab:clump}
\end{table*}

\subsection{Spatially Resolved Analysis of NIRCam Images}

We produce spatially resolved continuum maps in the rest-frame UV and optical using the NIRCam images.
We use the F200W images to probe the UV continuum powered by the recently-formed stars. This bandpass covers rest-frame 2000 -- 3000 \AA{} at the redshifts of our sample.
Two versions of the UV maps are created, one at original resolution in our fiducial reduction and the other matched to the same resolution of the F444W images.
The first version shows the  morphology at its highest resolution, while the latter is used to uniformly extract the NIRCam SEDs in subregions of the galaxy. 
The NIRCam images also allow for a view of the resolved morphology in the optical, which was inaccessible with the earlier \hst{} imaging.
The optical continuum is sensitive to light from  older stellar populations as well as the nebular continuum powered by young stars.
We carefully identify filters that are least impacted by strong emission lines (in particular, \hb{}, [\oiii{}], and \ha{}), based on their photometric redshifts and the \beagle{} SEDs.
The F356W images are used for 4 galaxies, F410M for another 5 galaxies, and F444W for the remaining 3 
sources. These filters sample the continuum emission at rest-frame wavelengths of 4000 -- 5300 \AA{}.

As already mentioned in Section 3.1, the galaxies in our sample power prominent rest-optical emission lines  ([\oiii{}]+\hb{}) which can heavily contaminate one or more NIRCam filters (F356W, F410M, or F444W).
This leads to large flux excesses in the contaminated filters as compared to adjacent continuum-dominated filters (e.g., F356W - F444W < -1 for EGSZ-9314453285 as shown in Table \ref{tab:sample}).
At the redshifts of our sample, the [\oiii{}]+\hb{} emission lines fall in the F356W filter for 8, and the F410M for 4 galaxies.
We estimate the continuum from the adjacent filters that are least contaminated by emission lines, which is then subtracted from the image of line-dominated filter. The resulting maps correspond to the distribution of the [\oiii{}]+\hb{} emission lines, which we will compare with the UV continuum morphology.

The high-resolution NIRCam images also allow us to characterize the structural properties for our sample, which are often resolved into multiple components.
We visually identify the UV clumps associated with each of galaxy from the NIRCam  F115W, F150W, and F200W images, treating those with a single component as one clump.
Following the procedures in \cite{Matthee2017}, we constrain the deconvolved half-light radius ($r_e$) in both UV and optical for individual clumps by fitting 2D asymmetric exponential profile with the {\tt IMFIT} software \citep{Erwin2015}.
We use the PSF constructed by stacking the star images in Section 2.1 for model convolution and estimate the size uncertainties via bootstrap resampling of 100 iterations.
For galaxies with multiple clumps, we first fit all clumps together with multi-component models to constrain their sizes.
When a satisfactory multi-component fit is not possible, we separate and fit the relatively fainter clumps alone, and refit the remaining clumps with multi-component models.
Objects with $r_e<$ 1 pixel in the rest-UV filters and 2 pixels in the rest-optical (roughly 1/2 FWHM of the corresponding PSF) are taken as unresolved, and we only list  their sizes as upper limits. 
The rest-UV images have a factor 2 higher resolution than the optical images, leading some of closely separated components resolved in the UV to be blended together on the optical images.
We thus place upper limits for their optical size based on the FWHM measured from the radial profile.
The resulting measurements are reported in Table \ref{tab:clump}, with the clumps ordered and labeled by their F200W magnitudes (see the next Section).

\subsection{Stellar population modeling}

The  spatial resolution and  wide wavelength coverage (from rest-UV to optical) of NIRCam enables us to constrain the stellar population properties of the individual structures within each galaxy.
In particular, the relative strength of rest-UV and optical continuum and the flux excess due to emission lines place much-improved constraints on the light-weighted stellar population ages and the dust content than was possible with UV data on its own. We derive the stellar population properties for the UV-selected clumps by fitting stellar population models to the PSF-matched NIRCam photometry. 
We will also consider the stellar population properties of the lower surface brightness emission that surrounds the clumps, testing 
whether the clumps are embedded in an older component of the galaxy.

The photometry for clumps and the surrounding medium have been extracted from PSF-homogenized NIRCam images, with all filters convolved to the same resolution of F444W.
The clump photometry is measured with a fixed 6-pix (0.18\arcsec) diameter aperture, which, at the typical UV sizes of these structures, is large enough to capture the emission from the central region but avoids contamination from the surrounding medium.
The aperture loss is taken into account based on the PSF-convolved models from our exponential profile fitting on the F200W images.
We also estimate the photometric errors with blank apertures, taking the standard deviation from at least 500 apertures randomly put on the 10\arcsec{}$\times$10\arcsec{} source-masked images.
The photometry for the lower surface brightness medium between the clumps is calculated by masking all pixels within 4 pixels from any clumps to reduce their contamination. 
To define the outer extent of this inter-clump medium, we use the galaxy segmentation maps produced with {\sc Photutils}, including the pixels that are 2$\sigma$ above the background.
The uncertainties are derived from the standard deviation of the background pixels in the same 10\arcsec{}$\times$10\arcsec{} region.

The NIRCam photometry is then fit with \beagle{} to derive constraints on the stellar population and dust properties in the different components.
The redshifts for each clump and the surrounding medium are fixed to those fitted with the integrated galaxy photometry (see Table \ref{tab:sample}), allowing us to focus on the SED differences induced by the variations of the stellar population ages or the dust content.
We adopt the same models and settings as those for constraining the integrated properties (see Section \ref{sec:sample}), also assuming a CSFH.
We report the inferred absolute UV magnitudes, stellar masses, light-weighted stellar population ages, V-band optical depths, and the rest-frame EW [\oiii{}]+\hb{} for each of the components in Table~\ref{tab:clump}.

\section{Results and Discussion}

\begin{figure}
    \centering
    \includegraphics[width=0.45\textwidth]{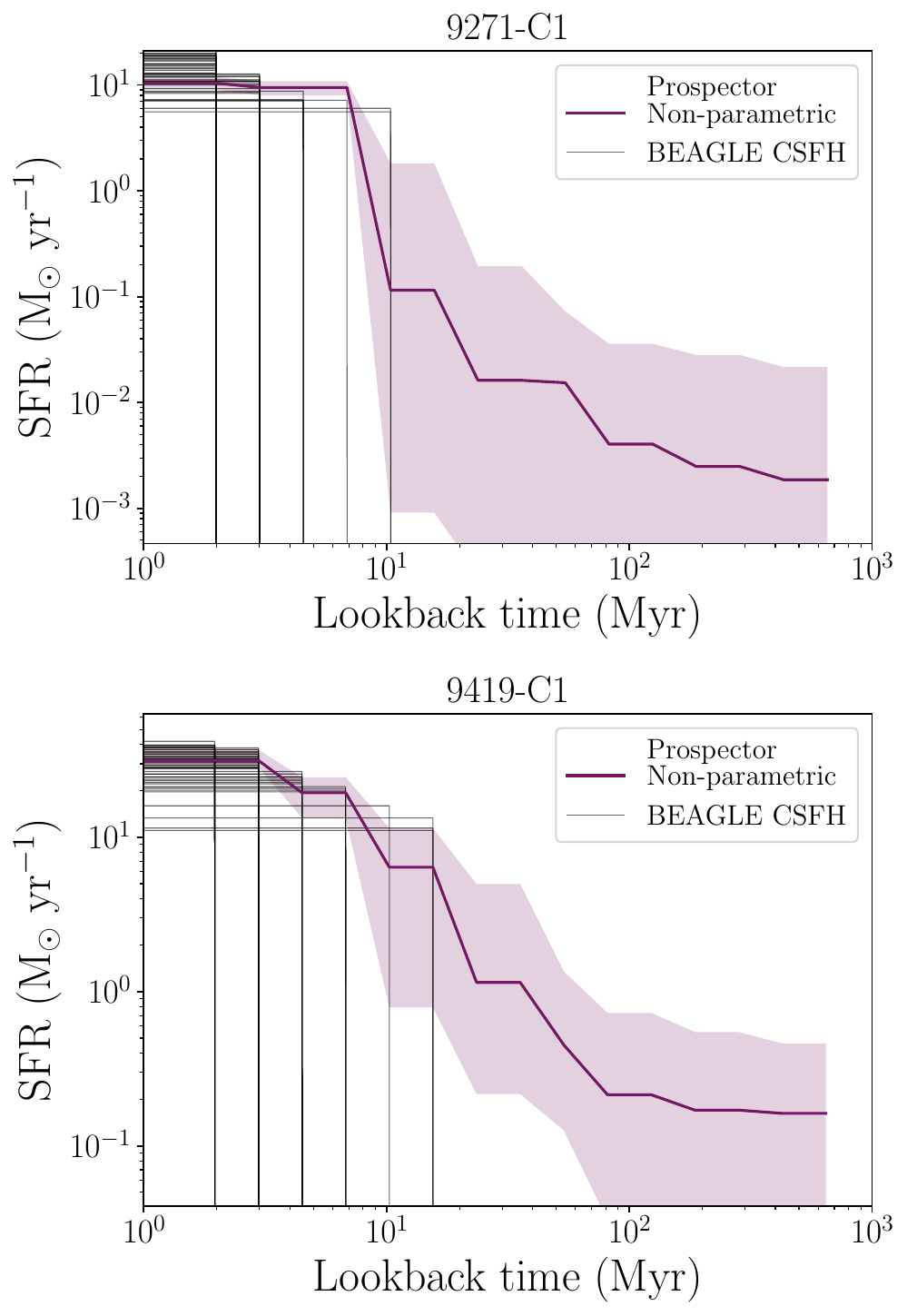}
    \caption{Non-praramteric SFH fits with {\sc Prospector} (magenta) to two extremely young clumps, 9271-C1 and 9419-C1, following methodology of \citet{Whitler2022}.
    Also shown in black are their \beagle{} CSFH fits (black), which favors extremely young light-weighted ages ($<$ 5 Myr).
    However, with the non-parametric SFH fits, it is clear that old stars with age $>$ 50 Myr could still be hidden in these systems, leading to a total stellar mass increasing by a factor of $\sim$ 5. Longer wavelength imaging 
    from MIRI will help clarify the presence of these older stars.}
    \label{fig:nonparam}
\end{figure}

\begin{figure*}
    \centering
    \includegraphics[width=1\textwidth]{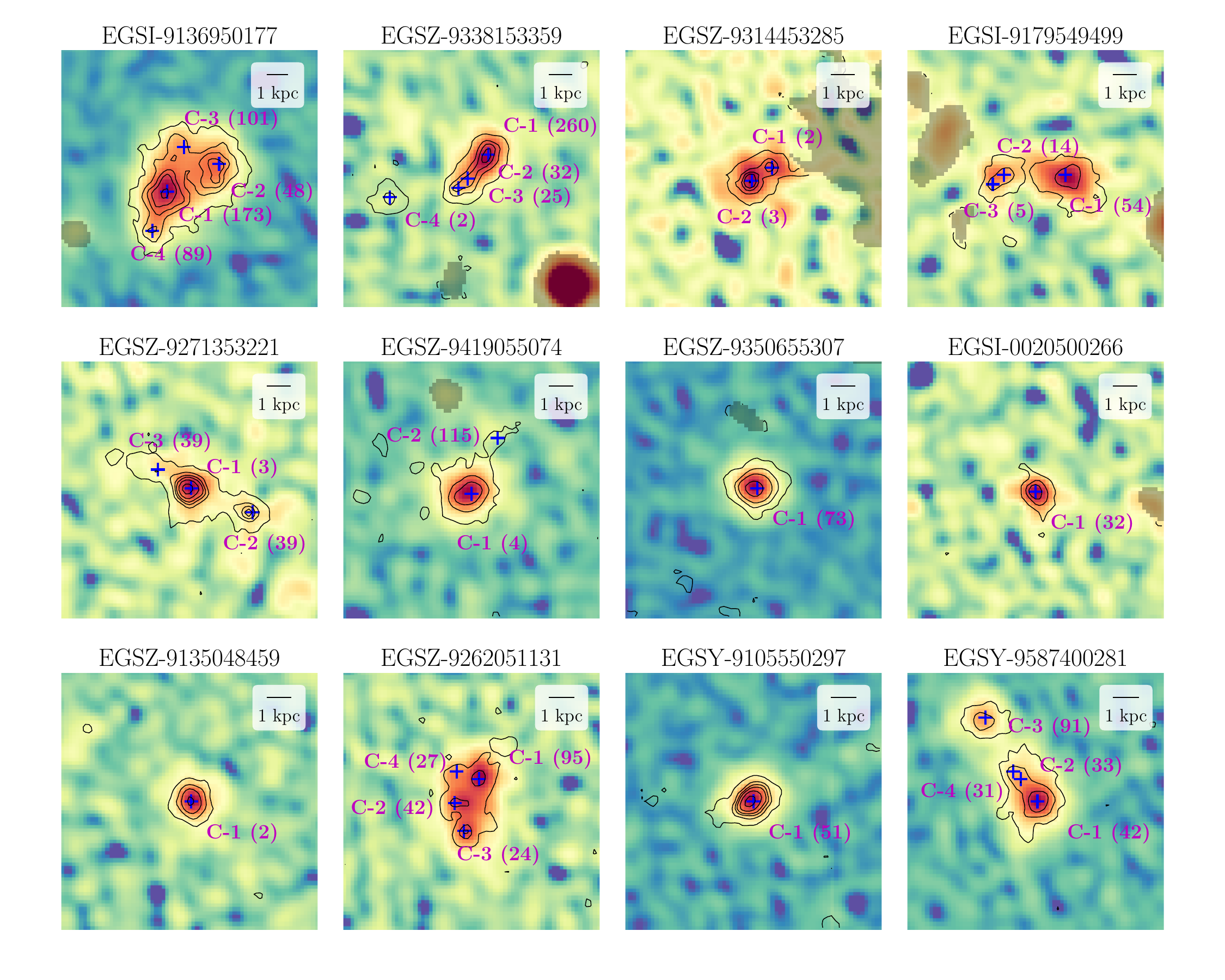}
    \caption{Comparison of the rest-optical continuum with the rest-frame UV continuum in 12 galaxies at $6<z<8$, constraining the relative distribution of old and young stars in reionization era galaxies.  The rest-optical continuum is shown 
    in color (with red denoting bright optical continuum), with rest-UV overlaid as contours. 
    The locations of the  clumps identified in the short-wavelength NIRCam photometry are marked with a blue cross, and the clump ID are also shown in magenta.
    We also note the stellar population age (assuming CSFH) in Myr for each clump in brackets right after the clump ID.
    Nearby contaminants have been shaded in gray. The rest-optical continuum is as clumpy as the rest-UV, with emission peaking in the young star forming complexes. 
    The images do not reveal optically-bright clumps that are not seen in the UV, consistent with the majority of mass being contained in the star forming complexes.}
    \label{fig:uv-opt}
\end{figure*}

\begin{figure*}
    \centering
    \includegraphics[width=1\textwidth]{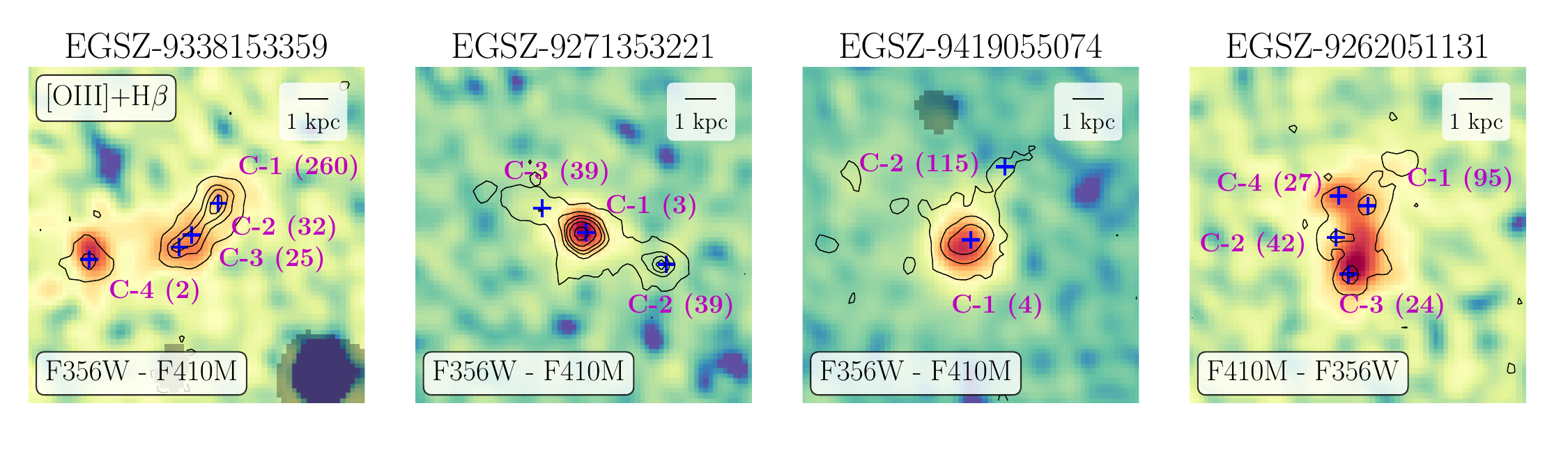}
    \caption{Spatial maps of 
    [\oiii{}]+\hb{} emission for 4 objects in our sample. UV continuum contours are overlaid.   Similar to Figure \ref{fig:uv-opt}, the locations of individual UV clumps are marked with a cross with the ID shown in magenta and the stellar population age (in Myr, assuming CSFH) in brackets.
     To create line maps, we use filters (denoted in bottom-left of each panel) that are dominated by strong emission lines (see Figure 2), and subtract the underlying optical continuum using adjacent filters with minimal line contributions (the second filter listed in the bottom-left).
     The [OIII]+H$\beta$ emission is found to vary considerably across 1 kpc scales in $z\gsim 6$ galaxies.
    }
    \label{fig:line_map}
\end{figure*}

In this section, we present our results on the spatially resolved properties of $z\sim 6-8$ galaxies based on the \jwst{}/NIRCam imaging. 
We begin by investigating the rest-UV structures in our galaxies (\S \ref{sec:res_clump}), focusing on the spatially-resolved  properties of the star forming clumps revealed in individual galaxies.
We then investigate the light distributions of both optical continuum and emission lines, and compare them with the location of the UV clumps (\S \ref{sec:res_opt}).
Lastly, we discuss the nature of the star forming clumps and their implications for mass assembly in $z>6$ galaxies (\S \ref{sec:res_size}).

\subsection{Star Forming Clumps in $z\sim 6-8$ Galaxies}\label{sec:res_clump}

Previous studies of $z\sim7$ galaxies have characterized the resolved structure in early galaxies using 
\hst{}/WFC3 imaging \citep[e.g.,][]{Ouchi2013,Sobral2015,Bowler2017,Matthee2017}, revealing important insights into the clumpy nature of 
luminous star forming systems. In the following, we take advantage 
of the higher spatial resolution, improved sensitivity, and additional filters of \jwst{}/NIRCam to build on these  studies investigating the internal properties in $z>6$ galaxies.

The NIRCam rest-UV images reveal a range of structures in our galaxies (Figure \ref{fig:rgb}).  The majority (8/12) of the sample are resolved into 2 -- 4 clumps, with four of these galaxies showing 4 clumps.
For the eight galaxies, the clumps are separated by $\approx$ 0.3 -- 4.3 kpc, with the closest separations found in  EGSZ-9338153359 (0.6 kpc), EGSI-9179549499 (0.6 kpc), and EGSY-9587400281 (0.3 kpc). 
The rest of the sample is well-described by a single component.  These basic properties are broadly similar to the previous findings derived from \hst{} imaging \citep[e.g.,][]{Ouchi2013,Sobral2015,Bowler2017,Matthee2017}.
The individual clumps can be quite luminous, with \muv{} spanning from -18.66 to -20.67. The clumps often dominate the total UV light of each galaxy, contributing at least 76\% of the total F200W flux.
The most luminous galaxies ($M_{\scriptscriptstyle{\rm UV}}<$-20.7) appear more likely to contain 3 -- 4 clumps, while the fainter ($M_{\scriptscriptstyle{\rm UV}}>$-20.7) systems tend to show only one single clump.
This suggests a multi-component structure is common in the most luminous galaxies with single clumps being common at lower luminosities, in agreement with  previous studies \citep[e.g.,][]{Bowler2017,Bouwens2022b}.
These high surface brightness UV structures are sites of vigorous star formation \citep[e.g.,][]{Dekel2014,Guo2015,Vanzella2017,Vanzella2019}

While these UV structures are often marginally resolved with the WFC3 images, the higher-resolution NIRCam data allow for improved constraints on their sizes.
The UV half-light radius (\ruv{}) of each clump has been constrained by profile-fitting with 2D exponential models. The derived values range from $<149$ pc up to 480 pc.
The small sizes are comparable to those of complexes of star clusters found at high redshift \citep[e.g.,][]{Vanzella2017,Vanzella2020,Bouwens2022b}, which we will discuss later.
The profile fits show that the UV structures are not always symmetric, with ellipticities ranging from 0.0 -- 0.7 when resolved.
The asymmetric morphology is clear in the images and can be a signature of interactions among the clumps or suggests that these structures are unresolved conglomerations of smaller unrelaxed structures. This is also consistent with the close separations among the clumps found in some systems, where two peaks are found at a distance of just $\approx$ 2 pixels ($\approx$ 300 pc) with both being very compact unresolved systems (e.g., EGSZ-9338153359, and EGSY-9587400281). Given their close separation, such clumps are likely interacting and will presumably merge together.

Prior to the {\it JWST} era, our interpretation of these clumpy structures was stunted by the absence of rest-optical imaging at comparable spatial resolution as provided by {\it HST}. The longer wavelength 
emission is critical for assessing the age of the clumpy structures, while also allowing dusty or old red structures (that may dominate the galaxy mass while being faint in the UV) to be identified. NIRCam provides  access to the rest-optical emission with some of its LW filters.
As described in \S2.4, we have fit the SEDs of all clumps identified in the images of our galaxy samples using the \beagle{} tool. 
The modeling shows that the stellar masses for the clumps vary from $8.6\times10^6$~$M_\odot$ to $5.6\times10^8$~$M_\odot$ (assuming CSFH). 
When combined with the estimated sizes of the clumps, this corresponds to a stellar mass surface density of $8$ to $\approx$~1117~$M_\odot$~pc$^{-2}$.
The SED fits also reveal variations of dust content across individual galaxies with multiple components, with the optical depths varying from negligible values up to 0.1 -- 0.3 (e.g., EGSI-9136950177, EGSZ-9271353221, and EGSZ-9419055074), consistent with gradients identified in ALMA far-IR continuum observations \citep[e.g.,][]{Bowler2022}.
The variation in UV color in different clumps is clear in 
Figure~\ref{fig:rgb}, with some star forming complexes being much redder in the UV than the others (e.g., those in EGSI-9136950177, and EGSZ-9271353221).

The ages of the clumpy star forming complexes give some insight into 
their nature. The inferred light-weighted ages (assuming a CSFH) are fairly young, with  a median age of 36 Myr. This is in 
general  younger than the star-forming clumps uncovered at lower redshifts \citep[e.g.,][]{Guo2015}. However, a wide range of ages can also be found within seven of the galaxies. In the case of EGSZ-9338153359, the youngest clump is dominated by an extremely 
young (2 Myr) stellar population, while another clump 2 kpc away appears much older ($>$200 Myr). It is clear that at least in the most 
luminous star forming galaxies at $z>6$, the integrated SED is 
comprised of the sum of several distinctly different stellar populations with varying light-weighted ages and dust content.

However as recently shown by \cite{Whitler2022}, even galaxies with light dominated by extremely young massive stars, a significant 
number of old stars can still be hidden.
By adopting the same approach as \cite{Whitler2022}, we fit the SEDs of the youngest star-forming clumps with non-parametric SFH, using the {\sc Prospector} \citep{Johnson2021} code (also see \citealt{Whitler2022b}). 
We adopt the continuity prior with a fixed redshift $z_{\rm form} = 20$ at which the star formation starts.
In doing so, we allow for significant amount of star formation at earlier epochs than CSFH models in cases where the galaxy has experienced a recent burst. 
In Figure \ref{fig:nonparam}, we present the resulting constraints on the past star formation with {\sc Prospector} for two of the clumps, i.e., 9419-C1, and 9135-C1.
The results suggest that stars older than 50 Myr could still contribute up to 25\% of the total stellar mass of the star forming complexes that have light-weighted CSFH ages younger than 5 Myr, increasing the derived stellar mass by a factor of a few times. In this case, the clumps ($>$150 pc in scale) identified in NIRCam would potentially be dominated by the light from very young recently-formed star clusters ($<$10 pc in scale and unresolved in our imaging), but would also contain fainter older clusters or field stars from previous generations of star formation. Such old stars are seen in nearby complexes with very young star clusters \citep[e.g.,][]{Aloisi2007,Corbin2008,Grebel2000}.

\subsection{Spatially-Resolved Maps of Optical Continuum and [\oiii{}]+\hb{} Emission}\label{sec:res_opt}

\begin{figure*}
    \centering
    \includegraphics[width=0.8\textwidth]{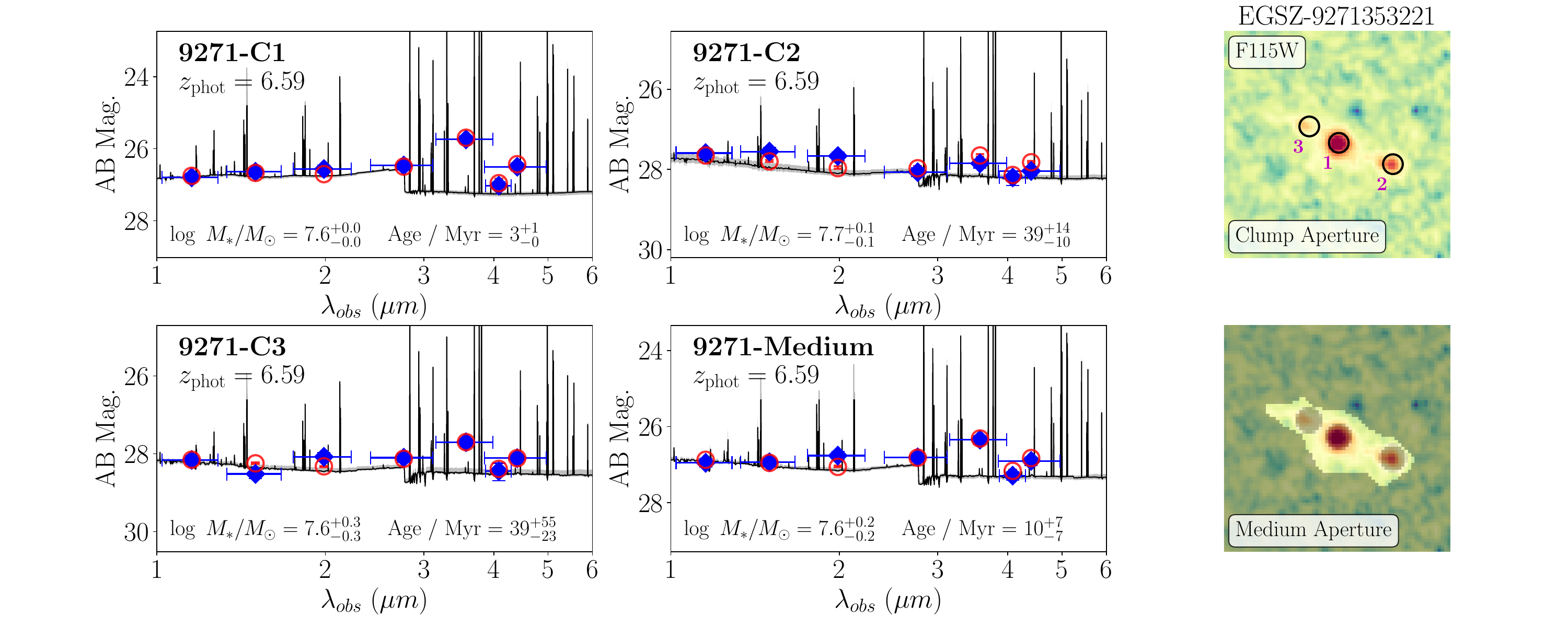}
    \caption{Resolved NIRCam SED fitting of individual clumps within a $z\simeq 6.59$ galaxy in our sample. The four left panels show the SEDs and \beagle{} model-fits, while the right panel shows the apertures used for the clumps and the 
    inter-clump medium.  The SEDs reveal clear light-weighted age differences in the different star complexes, clearly indicated by varying emission line excesses across the galaxy.}
    \label{fig:clump_sed}
\end{figure*}

To understand the variations of the clump properties in more detail,  we make spatially resolved maps of the rest-optical continuum, comparing them against the UV maps.
Based on the redshifts and the \beagle{} SED fits to the integrated photometry, we are able to identify filters that are least impacted by nebular lines and thus are dominated by the continuum emission from the nebular gas and older stellar populations.
We present the optical continuum morphology in Figure~\ref{fig:uv-opt}. In general, the optical morphology appears to also be clumpy, resembling the structures seen in the UV. We do not find evidence for very red clumps dominated by extremely old stellar population or obscured star formation that were previously hidden in the UV.  Such reddened nuclear structures may appear at later epochs or in higher mass reionization-era galaxies (also see the recent results from the GLASS-JWST program in \citealt{Treu2022}). 
The overall very similar rest-UV and rest-optical morphology for $z\gtrsim6$ galaxies is also consistent with the recent findings with the NIRCam imaging in the GLASS-JWST program \citep{Treu2022}.
For the systems in this sample, the rest-optical emission peaks where the UV clumps are located. For each clump, the offsets between the optical and UV peaks ($\lesssim$ 0.03\arcsec{}) are found to be within the astrometric uncertainties (0.015--0.025\arcsec{}). It is apparent in Figure~\ref{fig:uv-opt} that the ratio of UV to optical luminosity varies from clump to clump, with some of UV-bright regions significantly fainter at  longer wavelengths (e.g., 9271-C2), reflecting the variations in age and 
attenuation discussed above. This also leads to observable offsets between the integrated barycenter of the two maps (i.e., the first-moment flux weighted center; see \citealt{Bowler2017}), with the largest offset found in EGSZ-9338153359 at 0.16\arcsec{}.

The strength of [\oiii{}]+\hb{} emission lines tracks the locations of HII regions. The equivalent width  has been shown to strongly correlate with the age of the stellar population (and gas conditions) in the 
extreme emission line galaxy regime that is common at $z>6$ \citep[e.g.,][]{Tang2019,Endsley2021,deBarros2019}. Much more 
prominent line emission is seen in very young stellar populations with weak underlying optical continuum.
Taking advantage of the NIRCam LW filters that are heavily contaminated by these lines, we map the [\oiii{}]+\hb{} emission to provide additional insights into the distribution of the \hii{} regions and the age variations among clumps. 
We carefully isolate these filters by their clear flux excess relative to the continuum filters in the \beagle{} SEDs at the inferred redshift, and then subtract the underlying optical continuum emission (which we described above).
In Figure \ref{fig:line_map}, we present the emission line maps for four galaxies with large inferred variations of clump ages from our SED fits.
It is clear that the youngest UV structures (e.g., 9338-C2, and 9271-C1) are found where the line emission is strongest relative to the continuum.
The brightest clumps in the UV do not always power large flux excess in the line-dominated filters. We find this in particular occurs toward clumps 
with older stellar population ages (e.g., 95 -- 260 Myr for 9338-C1 and 9262-C1 assuming a CSFH). Because we find significant age variations in clumps across individual galaxies, we also find varying EW of the [\oiii{}]+\hb{} emission lines.  
This is also consistent with the recently studies that have reported variations of the FIR [OIII] 88 $\mu$m emission line properties that also traces the ionized gas in $z\simeq7$ galaxies \citep{Wong2022,Witstok2022,Akins2022}.
Measurements of the emission line EWs in the NIRCam photometry help to break the degeneracy between age and dust, improving constraints on both.
For example, in EGSZ-941905574, the brightest clump (9419-C1) has clearly red UV colors (F150W-F200W = 0.33 mag), but powers a prominent knot on the [\oiii{}]+\hb{} map.  The \beagle{} modeling favors a younger stellar population age ($4_{-1}^{+2}$ Myr) than the clump in this system with bluer UV colors but negligible emission line excesses ($115_{-58}^{+86}$ Myr). Many spectral properties (i.e, the [OIII]/[OII] ratio) correlate with the [\oiii{}]+\hb{} EW (e.g., \citealt{Tang2019,Du2020, Tang2021}).  The variations in [\oiii{}]+\hb{} EW across individual galaxies 
in our sample suggest that there are likely to be strong gradients in line ratios across different clumps in a given galaxy. Care must be taken to interpret future NIRSpec/MSA observations which may only be sensitive to one of several  clumps in an individual galaxy. The measured line ratios may not be representative for all bright star forming complexes within the galaxy.

While many clumps in our galaxies show very young light-weighted ages ($\lesssim$ 25 Myr), it is conceivable that they may be embedded in an older stellar population that is more smoothly distributed.
To test this, we constrain the stellar population properties of the lower surface brightness interclump medium by fitting its NIRCam SEDs with \beagle{} in the same way as that for the clumps.
As described in Section 2, the photometry for this region is extracted by summing the flux at 2$\sigma$ above the background, but masking the pixels (distance $<$ 4 pxiels) close to the location of any clumps.
We also consider different choices of masking radius (5 and 6 pixels), but we find consistent results on the inferred stellar population ages for the medium.
This suggests that our results are unlikely to be biased by contamination from the clump flux.
In Figure \ref{fig:clump_sed}, we present an example of resulting fits (EGSZ-9271353221), along with  the apertures used to extract the photometry.
In this galaxy, the interclump region also shows a  flux excess in the F356W filter compared to the adjacent continuum-dominated filter, with strength in between the three clumps.
While slightly older than the brightest clump (9271-C1, at $3_{-0}^{+1}$ Myr), it has a clearly smaller light-weighted age ($10_{-7}^{+7}$ Myr) than the other two clumps ($\approx 33$ Myr, all assuming a CSFH).
The light-weighted ages for the interclump medium still fall in the age range spanned by the clumps when we use different masking radius ($15 - 20$ Myr with masking radius 5 -- 6 pixels). 
When considering the full sample, our results suggest that the interclump medium is not always older than all of the clumps, with inferred ages spanning 2 -- 83 Myr.  When the lower surface brightness regions are found to older than 50 Myr, clumps with comparable ages can also be seen in the same system.  It is conceivable that such extended  and diffuse structures consist of stars stripped off from the clumps of similar ages during their interactions.

\subsection{The nature of compact star-forming clumps}\label{sec:res_size}

\begin{figure*}
    \centering
    \includegraphics[width=0.45\textwidth]{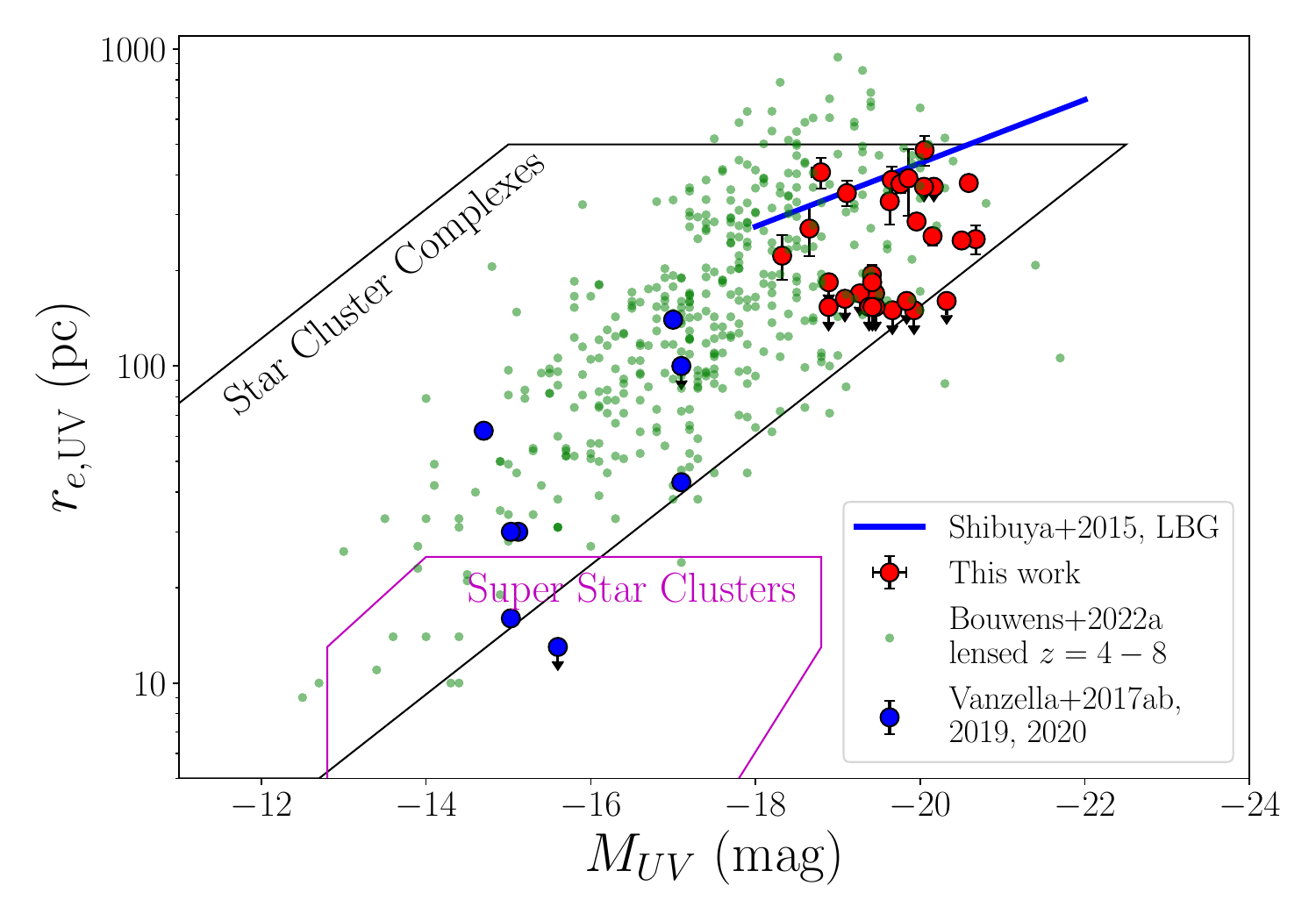}
    \includegraphics[width=0.45\textwidth]{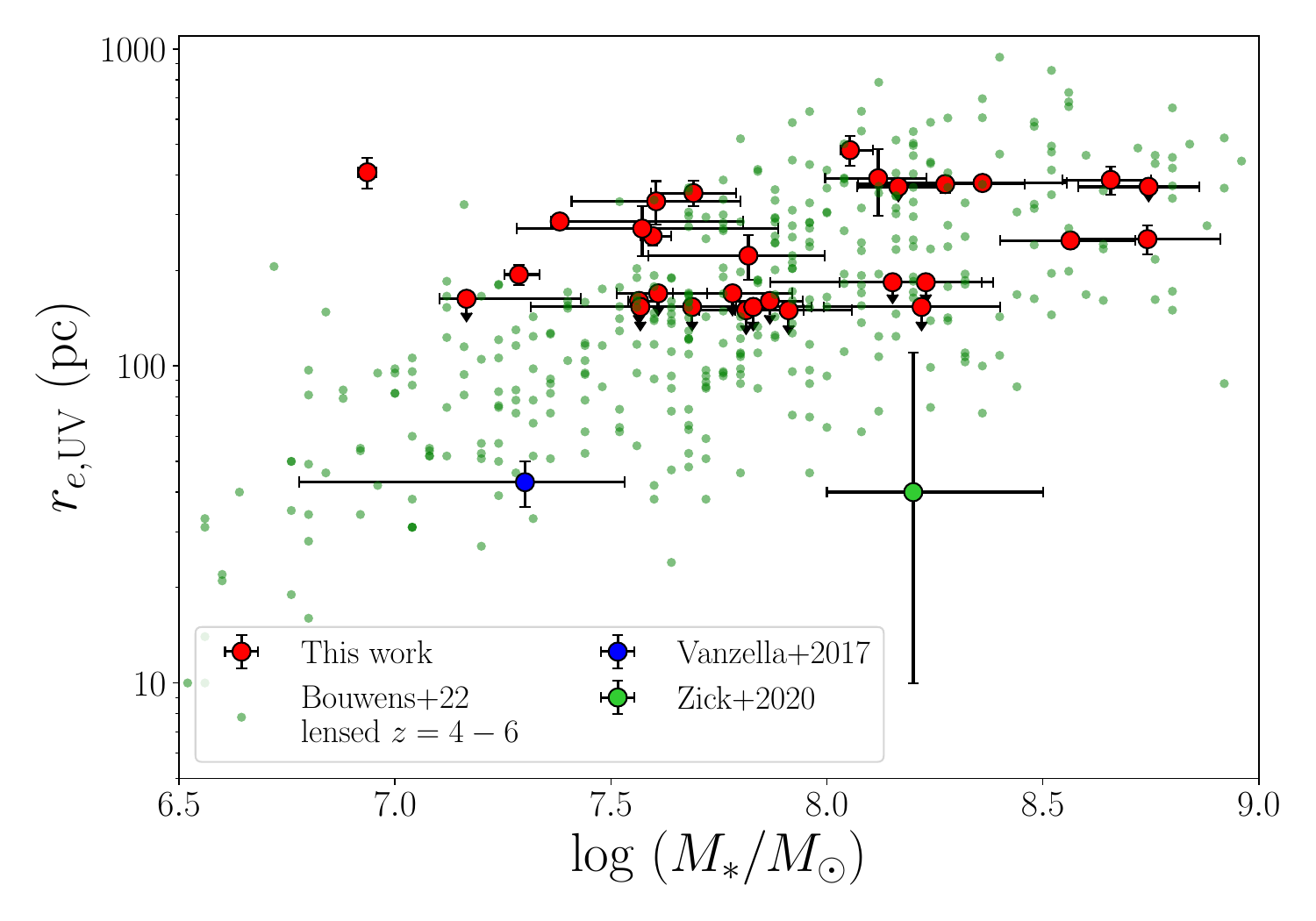}
    \caption{Relationship between size and UV luminosity (left panel) and size and stellar mass (right panel). 
    The star forming complexes identified in this work are denoted as red circles, with upper limits marking systems that are unresolved. 
    Following \citealt{Bouwens2021}, we also show in the left panel the regions occupied by star cluster complexes and super star clusters with polygons.
    The luminosity and size ranges spanned by these polygons match observations of star clusters at lower redshifts \citep[e.g.,][]{Bastian2006,Livermore2015,Zanella2019,Messa2019}.
    The clumps in the $6<z<8$ galaxies of this paper are consistent with the relation between size and luminosity derived from gravitationally lensed galaxies in the {\it Hubble} Frontier Fields, occupying the larger end of the relation \citep{Bouwens2022b}. These clumps correspond to star forming complexes, likely powered by multiple smaller super star clusters that are below our resolution limit. 
    We also show the sizes previously inferred for highly magnified stellar clumps in z = 2 -- 6 gravitationally lensed galaxies that are thought to be candidates for proto globular clusters \citep[e.g.,][]{Vanzella2017,Vanzella2017b,Zick2018,Vanzella2019,Vanzella2020}.
    It is clear that surveys using gravitational lensing with higher spatial resolution will be required to characterize these cluster-scale components. 
     }
    \label{fig:muv-re}
\end{figure*}

The clumpy star forming complexes identified in this paper resemble 
compact structures previously uncovered at $z\gtrsim6$ leveraging the magnification provided by gravitational lensing \citep[e.g.,][]{Kawamata2018,Bouwens2021,Bouwens2022b,Mestric2022}.
In Figure \ref{fig:muv-re} (left panel), we show the position of these clumps in the 
plane of rest-UV size and UV luminosity. 
For reference, we compare to magnification-corrected measurements of the  gravitationally-lensed galaxies identified in the {\it Hubble} Frontier Fields from \cite{Bouwens2022b}.  
Both the lensed and unlensed clumps follow a similar relationship between the luminosity and size, with the sources presented in this paper mostly probing the luminous end (given the absence of magnification).
The star forming clumps are not surprisingly (on average) smaller than individual Lyman Break Galaxies (LBGs) at these redshifts (adopting the \citealt{Bouwens2022b} linear fit of the luminosity-size relation for the \citealt{Shibuya2015} LBGs), occupying a region more similar to the star cluster complexes found at lower redshifts \citep[e.g.,][]{Bastian2006,Livermore2015,Zanella2019,Messa2019}.
The resolution limit of NIRCam at these redshift prevents us from identifying  substructures within the $\approx$150 pc scales spanned by many of our clumps.
It is likely that several  young star clusters ($\simeq 10$ pc in size) are dominating the light from our clumps. Characterizing 
such smaller structures requires gravitational lensing  \citep[e.g.][]{Vanzella2017,Vanzella2017b,Vanzella2019,Vanzella2020}. NIRCam imaging of 
massive galaxy clusters will allow many such star clusters to be identified, complementing the characterization of larger clump complexes that is possible in the field.

The internal structure of $z\gsim 6$ galaxies, with most of the stellar mass locked in bright star forming complexes, points to a rapid phase of mass assembly, consistent with the large sSFR expected at these high redshifts (e.g., \citealt{Topping2022, Stefanon2022, Leethochawalit2022}). The presence of 
multiple clumpy components separated by $<$1 to 4 kpc (and asymmetric single 
components) suggests merger activity may be common, as suggested by 
pair counts \citep{Duncan2019}.
In the right panel of Figure \ref{fig:muv-re}, we show the mass-size 
relationship of the star forming complexes, as compared to the 
faint lensed galaxies from \cite{Bouwens2022b}.  We note that the lensed 
galaxies do not have rest-optical measurements, so the stellar masses have 
been converted assuming a fixed stellar age, consistent with what has been adopted in (\citealt{Bouwens2022b}; also see \citealt{Kikuchihara2020}).
The position of the UV-bright clumps on the size - stellar mass plane implies large stellar mass surface densities, ranging from 8 -- 1117 $M_\odot$ pc$^{-2}$.
The largest values approach those measured for the most compact star-forming complexes uncovered with lensing  \citep[e.g.,][]{Vanzella2017,Zick2018}, and are potentially  sites of proto-globular clusters in formation  \citep[e.g.,][]{D'Ercole2008}.
The assembly and merging of these dense structures may dominate the mass growth of the galaxy \citep[e.g.,][]{Ouchi2013,Conselice2014}, helping eventually to build early bulge-like structures at high redshifts.

\section{Summary}

We characterize the internal structure of 12 star forming 
galaxies at $z\simeq 6-8$ using {\it JWST}/NIRCam imaging 
of the EGS field. The galaxies are bright (F200W=24.9 to 26.6) 
and were selected in {\it HST} imaging of the EGS field. 
The NIRCam images sample the SEDs from the rest-frame UV to optical with up to 150 pc resolution, enabling maps to be made of the relative 
distribution of young star forming regions and old 
stellar populations.  We also create maps of [OIII]+H$\beta$ emission using filters dominated by strong emission lines, tracking the locations of the HII 
regions.  We characterize the size distribution and SED-based properties of the sub-components identified in each galaxy.
We summarize our primary conclusions below.

1. The $z\gsim 6$ galaxies generally appear clumpy in the rest-UV, with the light dominated by  star forming complexes (or ``clumps'') with stellar masses of $9\times$10$^{6}$ M$_\odot$ to $7\times$10$^{8}$ M$_\odot$. The UV-detected clumps are generally very compact in the NIRCam images with median effective radii of 236 pc (ranging from $< 150$ pc to 480 pc). Most of the galaxies in our sample sample (8 of 12) are comprised of multiple bright clumps with typical separations of 0.3 to 4.3 kpc, similar to what has been seen in UV-luminous galaxies at $z\simeq 7$ (e.g., \citealt{Bowler2017,Matthee2017}). The bright star forming complexes are embedded in a more diffuse  (``interclump'') UV-emitting stellar population. In our sample, we find that the clumps comprise on average at least 76\% of the UV luminosity. The 
star forming complexes vary significantly in their UV continuum colors, with variations seen within individual galaxies (as also seen at {\it HST} resolution by 
\citealt{Bowler2022}).

2. We characterize the ages of the star forming complexes using the SEDs of the clumps identified in the NIRCam images.  The light-weighted ages of the clumps in $z\gsim 6$ galaxies tend to be  young, with a median of 36 Myr (here assuming CSFH), much younger than the clumps seen in galaxies at $z\simeq 2$. Within a given galaxy, the clumps can vary significantly in age, with some complexes extremely young ($<$5 Myr) and others several kpc away showing much older stellar populations ($>$100 Myr). When we adopt more flexible star formation histories we find that the youngest 
clumps ($\lsim 5$ Myr) can be fit with non-negligible contributions from older stellar populations (e.g., \citealt{Whitler2022,Whitler2022b,Tacchella2022}). It is plausible that 
these very young clumps ($\simeq 150$ pc in size) are dominated by the light from a small recently-formed super star cluster which outshines the light of older stars 
in its vicinity.

3. The longer wavelength NIRCam imaging (2-5$\mu$m) allows us to test whether the young UV-emitting star forming complexes surround a central nuclear structure with older stars which dominates the mass. In the 12 $z\gsim 6$ galaxies considered in this paper, we find that the distribution of rest-optical continuum closely matches 
that seen in the rest-UV, with emission peaking in the bright star forming complexes. We find no evidence in the rest-optical for old and red central clumps which are undetected (or faint) in the rest-frame UV. Such old nuclear structures may not yet be present in the brightest reionization-era galaxies. Instead a significant fraction of the galaxy mass appears to be situated in the clumpy star forming complexes. 

4. We  characterize the properties of the regions between the clumps, testing whether the star forming complexes are embedded in an older and more smoothly-distributed stellar population. In some cases, we do find that the interclump medium is 
dominated by an old ($\gsim 100$ Myr) stellar population which contributes a 
significant amount of mass to the galaxy. However in these galaxies, there is 
nearly always at least one UV-bright clump with a similarly old age as the more diffuse lower surface brightness population. The data therefore do not suggest 
the UV-bright clumps are situated in a uniformly older population of stars. 

5. We create maps of [OIII]+H$\beta$ emission using (continuum-subtracted) line-dominated filters in 
the LW NIRCam channels, enabling the distribution of HII regions to be mapped 
across the galaxies. We find significant gradients in the [OIII]+H$\beta$ 
strength across individual galaxies, with some star forming complexes 
powering much stronger line emission (for similar optical continuum strength) 
than other complexes located $\simeq 1$ kpc away. These variations are likely 
driven in part by the differences in the stellar population ages of the 
underlying star forming components. Given the correlation between rest-optical spectral properties and [OIII] EW (e.g., \citealt{Tang2019}), we expect significant 
variations in line ratios across UV-bright reionization era galaxies. Care must be 
take in interpreting results from NIRSpec MSA observations which may only probe a 
small sub-region within these systems.

6. The UV sizes and luminosities  of the star forming complexes in our sample are 
consistent with the relationship between effective radii and M$_{\rm{UV}}$ derived 
for faint gravitationally lensed galaxies in the Hubble Frontier Fields \citep{Bouwens2022b}, sampling the higher mass end probed by the lensing fields. The 
light from the clumps are likely powered by one or more super star clusters below our $\simeq 150$ pc resolution limit, perhaps providing signposts of proto-globular clusters in formation.

With most of their stellar mass found to be in luminous star forming complexes $\simeq$150-480 pc in size, reionization-era galaxies are clearly in a phase of rapid assembly. These structures may eventually merge together (e.g., \citealt{Ouchi2013, Conselice2014}), building denser nuclear structures. 
Future observations with {\it JWST} will soon build on this picture.
NIRSpec integral field spectroscopy targeting bright $z\gsim 6$ galaxies will soon reveal 
the kinematics and dynamical mass of the individual star forming components, while longer wavelength 
imaging with MIRI will help clarify the contribution of old stellar populations to the
youngest star forming complexes.
Sub-parsec resolution cosmological simulations will also complement our understanding of the conditions for forming such star forming complexes in early galaxies \citep[e.g.,][]{Calura2022}.
Collectively 
these studies will help establish the origin of the bright star forming complexes in $z\gsim 6$ galaxies  and their role in the build-up of early galaxy structures.

\section*{Acknowledgements}

The authors thank the anonymous referee for their helpful comments that improved and strengthened this paper.
DPS acknowledges support from the National Science Foundation through the grant AST-2109066.
RE acknowledges funding from NASA JWST/NIRCam contract to the University of Arizona, NAS5-02015.
LW acknowledges support from the National Science Foundation Graduate Research Fellowship under Grant No. DGE-2137419.

The authors thank Jacopo Chevallard for use of the \beagle{} tool used for 
much of our SED fitting analysis. We thank Gabe Brammer for providing the optical imaging of the EGS as part of CHArGE program.
This material is based in part upon High Performance Computing
(HPC) resources supported by the University of Arizona TRIF, UITS,
and Research, Innovation, and Impact (RII) and maintained by the
UArizona Research Technologies department.

\section*{DATA AVAILABILITY}
The JWST and HST imaging data used in this work are
available through the MAST Portal (\url{https://mast.stsci.edu/portal/Mashup/Clients/Mast/Portal.html}).
The photometry and analysis code used in this work will be shared upon reasonable request to the corresponding author.



\bibliographystyle{mnras}
\bibliography{main}



\appendix

\bsp	
\label{lastpage}
\end{document}